\documentclass[traditabstract]{aa}

\usepackage{array}
\usepackage{natbib}
\usepackage{amssymb}
\usepackage{bm}
\usepackage{subfigure}
\usepackage{fourier}
\usepackage{multirow}
\usepackage{graphicx}
\usepackage{rotating}
\usepackage{lscape}
\usepackage[english]{babel}															
\usepackage{enumerate}
\usepackage{url}

\usepackage{ulem}
\usepackage[usenames,dvipsnames]{xcolor}

\begin{document}

\title{Automated novelty detection in the WISE survey \\with one-class support vector machines}
\titlerunning{Automated novelty detection in WISE with OCSVM}

\author{A.~Solarz\inst{1}
\and M.~Bilicki\inst{2,1,3}
\and M.~Gromadzki\inst{4}
\and A.~Pollo\inst{1,5}
\and A.~Durkalec\inst{1}
\and M.~Wypych\inst{5}
}

\institute{ National Center for Nuclear Research, ul.\ Ho\.{z}a 69, 00-681 Warsaw, Poland \\
         \email{aleksandra.solarz@ncbj.gov.pl (AS); bilicki@strw.leidenuniv.nl (MB)}
         \and
Leiden Observatory, Leiden University, the Netherlands
        \and
         Janusz Gil Institute of Astronomy, University of Zielona G\'ora, Poland
\and Warsaw University Astronomical Observatory, Warszawa, Poland
 \and The Astronomical Observatory of the Jagiellonian University, Poland
}
\date{Received <date>/ Accepted <date>}
\abstract{
Wide-angle photometric surveys of previously uncharted sky areas or wavelength regimes will always bring in unexpected sources -- {novelties}
or even {anomalies} -- whose existence and properties cannot be easily predicted from earlier observations.
Such objects can be efficiently located   with novelty detection algorithms.
Here we present an application of such a method, called one-class support vector machines (OCSVM), to search for anomalous patterns among sources preselected from the mid-infrared AllWISE catalogue covering the whole sky.
To create a model of expected data we train the algorithm on a set of objects with spectroscopic identifications from the SDSS DR13 database, present also in AllWISE. The OCSVM method detects as anomalous those sources whose patterns -- WISE photometric measurements in this case -- are inconsistent with the model. 
Among the detected anomalies we find artefacts, such as objects with spurious photometry due to blending, but more importantly also real sources of genuine astrophysical interest.
Among the latter, OCSVM has identified a sample of heavily reddened AGN/quasar candidates distributed uniformly over the sky and in a large part absent from other WISE-based AGN catalogues. It also allowed us to find a specific group of  sources of mixed types, mostly stars and compact galaxies.  By combining the semi-supervised OCSVM algorithm with standard classification methods it will be possible to improve the latter by accounting for sources which are not present in the training sample, but are otherwise well-represented in the target set. 
Anomaly detection adds flexibility to automated source separation procedures and helps verify the reliability and representativeness of the training samples.  It should be thus considered as an essential step in supervised classification schemes to ensure completeness and purity of produced catalogues. }


\maketitle 

\section{Introduction}

Catalogues of astronomical objects derived from sky surveys serve as a foundation for any subsequent scientific analysis. 
 One of their primary uses is to provide information about statistical properties and spatial distribution of the observed sources, and to identify rare objects, especially those whose presence in the  dataset is not expected. 
Regardless of the aim of the survey, it is important to identify what characteristic properties each class of objects exhibits. This  information is crucial in order to separate the desired type of sources from the heap of collected data for further analysis.

The nature of an astronomical object can be determined most reliably by analysing its electromagnetic spectrum. However, even the largest spectroscopic surveys undertaken today,
 designed to provide  detailed information about each observed object, usually cover just a fraction of all the sources available for a given instrument.  
Photometric observations, on the other hand, are capable of delivering data for many more sources at a significantly faster rate and lower costs.

For photometric data the traditional tool for object separation are colour-colour (CC) and colour-magnitude (CM) diagrams, where various types of objects (like stars and galaxies) appear in separate areas due to differences in observed colours \citep[e.g.][]{walker89,pollo10,jarrett17}.
Today's largest photometric datasets, such as SuperCOSMOS \citep{scosmos}, WISE \citep{wright10}, or Pan-STARRS \citep{chambers}, contain  of the order of a billion catalogued sources each, which means that even now the traditional ways of dealing with the resulting catalogues by direct human inspection are not practicable.
These numbers are expected to grow by orders of magnitude with future experiments such as the LSST\footnote{\url{www.lsst.org}} or SKA\footnote{\url{www.skatelescope.org}}, so it is crucial to develop automated methods for source classification in the associated data products.
With the advent of self-learning algorithms, the task of source separation can now be dealt with  much more efficiently and  much more reliably,  due to the ability of the algorithm to work in a multidimensional rather than two-dimensional parameter space (e.g. CC or CM diagrams) as is usually the case for human analysis. 
  
Machine learning schemes are now widely used to automatically classify astronomical sources.  Owing to automated algorithms, selecting objects of significantly different properties (compactness, colour, etc.) from sky surveys has become 
quite straightforward \citep{zhang04,solarz2012,cavuoti14,shi15,heinis2016}. 
 However, depending on the nature of the survey, we can expect different types of objects to appear within the field of view. 
In surveys covering large areas of the sky and reaching deep enough to encompass significant amounts of both Galactic and extragalactic sources, the source separation is usually complicated. In such a case restricting the search to just a few  basic classes (e.g. stars, galaxies, and quasars) is not sufficient any more, as the closer we get to the Galactic plane the more diverse objects we can expect to find,  planetary nebulae, special types of stars (embedded in envelopes or undergoing catastrophic events), regions of interstellar matter, etc.
Moreover, the wavelength regime in which observations are made determines what kind of objects can be expected to appear. For instance, in optical surveys we will find far fewer dust-rich objects than in infrared (IR) ones, while hot stars clearly visible in optical and ultraviolet bands will fade away at longer wavelengths.
 On the other hand, if any unknown objects are present within the data, their properties should  stand out from the crowd of the expected ones. 
However, detecting these outliers is not straightforward as it is not uncommon for rare objects to mimic the appearance of the well-known ones; for instance,  a star  and a compact galaxy could both be classified as the same source type  based on their angular size only. 

 In \citet{kurcz16} an attempt was made to perform an automated, supervised  source classification of IR sources from the all-sky survey conducted by the WISE satellite. The training sample was based on the most secure identifications of SDSS spectroscopic sources divided into three classes of expected (normal) objects: stars, galaxies, and quasars. However, such a standard approach of supervised classification is not designed to correctly handle objects with patterns absent during training, or in other words, anomalous sources.
Moreover, especially in low-resolution surveys like WISE, in areas of high observed source density such as low Galactic latitudes, measurements are plagued by effects of overcrowding and therefore blending of objects. In such cases the measured properties of objects can display deviant characteristics and create further training biases.
These issues were partly avoided in \cite{kurcz16} by removing the data from the lowest Galactic latitudes ($|b|<10^\circ$ and wider by the Galactic bulge) in order to improve the classification. Nevertheless, both the catalogue of galaxy candidates and especially of the putative quasars obtained there exhibited large-scale on-sky variations resulting from issues with the data themselves, but also -- and maybe more  importantly -- from imperfections of the classification approach.  One of the goals of the present study is to examine whether those results could be explained by the existence of unaccounted for anomalous sources in the WISE data, and what the prospects are for improving that classification.

 By definition, characteristics of the unexpected sources are not known a priori, rendering the standard multi-class approach inapplicable.
Novelty detection schemes offer a solution to these problems, and such methods are designed to recognise cases when a special population of data points differ in some aspects from the data which are  used to train the machine learning algorithm.
It is common to apply these methods to datasets which contain a large number of examples representing the `normal' populations, but for which data describing the `anomalous' populations are insufficient.
In this work objects inconsistent with the training data will be defined as anomalous, as in principle they should display novel/outlying properties in the parameter space. In other words, `unknown' patterns in the target set will manifest themselves in the form of points deviating from the `known' sources.
A comprehensive review of this type of methodologies developed for machine learning can be found in \citet{hodge04,agyemang06}; and \citet{chandola09}.
According to \citet{hodge04}, a user can approach the problem of novelty detection in three different ways.
The first  is based on {unsupervised clustering}, where outliers are detected without any previous knowledge about the data. The second approach uses 
 supervised classification, where data have to be prelabelled as normal and unknown. The third method is a mixture of the first  two  and is referred to as {semi-supervised recognition}, where the algorithm models the normality of the data, and no knowledge about the true nature of test data is assumed.
In this third approach the observer designs an algorithm to create a model of how normal 
 data behave based on a large number of representative examples introduced during training. 
Next,  the algorithm investigates previously unseen patterns by comparing them to the model of normality and searches for a score of {novelty}. This decision threshold is then used to infer whether the data are behaving in a different manner with respect to the training set or not.

A wide selection of outlier detection algorithms is currently available. Some, based on unsupervised approaches like  random forest (\citealt{rf98}), were used by \citet{baron16} to find SDSS galaxies with abnormal spectra. Other methods, based on semi-supervised graph-based methods like label propagation and label spreading
(described in detail in \citealt{chapelle06}), were used by \citet[][]{skoda16,skoda162}
to identify artefacts and interesting celestial objects in the LAMOST survey. 

In the present work we use a knowledge-based novelty detection method designed to create a boundary
  within the structure of the training dataset,  support vector machines (SVM, \citealt{vapnik}),  to show the power of anomaly detection algorithms and discuss how they could  improve automatic selection schemes for present and forthcoming surveys covering large areas of the sky. As a case study, we will search for potential new source classes in the WISE dataset. 
 
 The paper is organized as follows: in Sect.\ \ref{sec:Data} we present the data and the parameter space used by the SVM algorithm; a description of the  one-class SVM algorithm we use can be found in Sect.\ \ref{sec:Novelty} where we discuss the steps of anomaly detection and describe the training process;  Sect.\ \ref{sec:Results} contains the results of the application of those procedures to AllWISE data; a summary and conclusions are given in Sect.\ \ref{sec:Summary}.

\section{Data}
\label{sec:Data}
In this paper we perform  anomaly detection in the Wide-field Infrared Survey Explorer (WISE) data, aiming at improving the early all-sky classification results of \cite{kurcz16} and at searching for deviant objects with unexpected properties.
Exploration of the publicly available AllWISE catalogue  \citep{cutri13}, which contains over 747 million sources with photometric information, allows us to test the power of basic artificial intelligence algorithms for anomaly detection in order to obtain information about special objects contained within the dataset. 
Currently AllWISE is the deepest all-sky dataset available to the public which at the same time provides vast amounts of data that  can be used to test automatic schemes of classification and detection of novelty.

 The WISE telescope \citep{wright10}, launched by NASA in December 2009, has been scanning the whole sky, originally in four passbands ($W1$, $W2$, $W3$, and $W4$) covering near- and mid-IR wavelengths centred at 3.4, 4.6, 12, and 23 $\mu$m, respectively. 
The AllWISE Source Catalogue was produced by combining the WISE single-exposure images from the WISE 4-Band Cryo, 3-Band Cryo, and NEOWISE Post-Cryo survey phases (\citealt{mainzer14}).
The angular resolution of the filters is $6.1"$, $6.4"$, $6.5"$, and $12.0"$, respectively, and the sensitivity to point sources at the $5\sigma$ detection limit is estimated to be no less than 0.054, 0.071, 0.73, and 5 mJy, which is equivalent to 16.6, 15.6, 11.3, and 8.0 Vega mag, respectively\footnote{\url{http://wise2.ipac.caltech.edu/docs/release/allwise/expsup/sec2_3a.html}}.
 
\subsection{Source preselection: WISE $\times$ SDSS cross-match}
\label{sourcepreselection}
The source preselection and parameter space for the purpose of this study follows that of \cite{kurcz16}. Namely, we focus on reliable measurements only  in the $W1$ and $W2$ channels to maximise the completeness and uniformity of the sample.
We thus use AllWISE sources which meet  the following criteria:  profile-fit measurement signal-to-noise ratios $\mathtt{w1snr}\ge 5$ and $\mathtt{w2snr}\ge 2$; saturated pixel fractions $\mathtt{w1sat}$ and $\mathtt{w2sat}\le 0.1$. To ensure that we do not preserve any severe artefacts we also apply $\mathtt{cc\_flags}[1,2]\ne '\mathrm{DPHO}'$, which excludes sources with diffraction spikes, persistence, halos, or optical ghosts. We emphasise that we do not use the $W3$ and $W4$ channels for the preselection nor for the SVM analysis; the former band will only be employed in the verification  phase for CC plots. 

In the domain of knowledge-based machine learning it is necessary to create a template for the classification of known objects. Therefore, the training data should representatively sample  the underlying distribution of target objects within a given parameter space. In the case of this study, an ideal training set would be constructed from a subsample of securely measured WISE sources with well-defined types.
However,  since at present such datasets are not available for WISE, to create the basis for the training process an external dataset containing sources of interest is needed.
For this purpose we construct the training set by cross-matching the AllWISE dataset with the Sloan Digital Sky Survey (SDSS, \citealt{york00}) DR13 \citep{sdssdr13}, which provides spectroscopic measurements \citep{bolton12}. The SDSS spectroscopic sample includes over 4.4 million sources, where galaxies comprise 59\%, quasars 23\%, and stars 18\%. The cross-matching procedure was performed using a $1"$ matching radius and resulted in 3 million common sources, of which galaxies constitute 70\%, quasars 12\%, and stars 18\%. This sample of AllWISE sources with a counterpart in SDSS DR13 spectroscopic will  henceforth be referred to as AllWISE$\times$SDSS, and below we provide details on cuts applied to it before the training procedure.
\subsection{Parameter space}
As in \cite{kurcz16} where SDSS DR10 was used, the cross-match between AllWISE and SDSS DR13 practically does not provide galaxies fainter than Vega $W1=16$ or $W2=16$. For the sake of completeness we thus trim all our catalogues, including the target AllWISE, at these limits; the same applies to any other cuts described below. At the bright end the matched catalogue contains practically only stars, we thus apply additional criteria of $W1>9.5$ and $W2>9.5$ as otherwise such a population of bright stars would be identified as anomalies by our scheme. 

In the earlier related studies by \cite{kurcz16} and \cite{krakowski16}, where a more classical approach to supervised learning was applied, the SDSS-based training sets were purified of sources with problematic redshift measurements according to SDSS parameters such as \texttt{zWarning} and \texttt{zErr}; this was done to avoid type misidentifications which could have detrimental effects on multi-class source identification.  In the case of outlier detection schemes the aim is to search for sources which do not exhibit patterns learned during training. As such algorithms are not designed to provide distinctions between specific classes but rather to show unexpected sources, the quality of the redshift measurement is of little interest here. Even though the SDSS spectroscopic data themselves may contain anomalous sources, the focus of our study is to find interesting objects within the infrared WISE catalogue. Therefore, even if an objects exhibits deviant properties at optical wavelengths but is otherwise well-detected, it will still be included in our training sample as a known source.  
For that reason we do not apply any data cleaning on the SDSS spectroscopic database.

As mentioned above, our parameter space was limited to two out of four available WISE passbands: $W1$ and $W2$. This is to ensure as many objects as possible in the final catalogue as the $W3$ and especially the $W4$ filters have much lower sensitivities and a much shorter data acquisition period (limited to the cryogenic phase) 
which leads to their much lower detection rates than at the shorter wavelengths. The $W3$ and $W4$ passbands are dominated by upper limits and non-detections in the WISE database and  using them in our study would lead to  losing the majority of objects and introducing severely non-uniform distribution of the sources, and significant biases in the photometry.

To ensure the maximum coverage of the parameter space by known sources, instead of using the $W1$ and $W2$ measurements separately, we employ the $W1$ magnitude and the $W1-W2$ colour. Even though using flux measurements from each filter separately is mathematically equivalent to employing the colours derived from them (e.g. \citealt{wolf01}), usage of colours can enhance the spread area within the parameter space for the considered objects. 
Finally, to extend the parameter space, we also use a {concentration parameter} defined as the difference between flux measurements in two circular apertures in the $W1$ passband in radii equal to $5.5"$ and $11.0"$ centred on a source:
\begin{equation}
\mathtt{w1mag13} \equiv \mathtt{w1mag\_1} - \mathtt{w1mag\_3}\;,
\end{equation}
used previously by \cite{2MPZ,bilicki16,kurcz16}; and \cite{krakowski16}. It serves as a proxy for morphological information: extended sources will typically have larger \texttt{w1mag13} values than point-like ones (see Fig.~\ref{fig:w13}).
We emphasise that currently the WISE database does not provide any reliable extended source identifications nor isophotal magnitudes, except for a small subset ($\sim 500,000$) of objects in common with the 2MASS Extended Source Catalogue and for those in some of the GAMA fields \citep{cluver14,jarrett17}. 

To summarize this part, the chosen parameter space has three dimensions:
\begin{enumerate}
\item $W1$ magnitude measurement;
\item $W1-W2$ colour;
\item concentration parameter $\mathtt{w1mag13}$.
\end{enumerate}
All the WISE magnitudes will be given in the Vega system.
\subsection{Quality cuts}
To purify the data further, we apply two cuts on the concentration parameter. First, we require that $\mathtt{w1mag13} \geq 0$ to remove objects with measured flux decreasing with increased aperture, which are most likely artefacts of source extraction in high-density areas. This cut removes 3,360 sources from the AllWISE$\times$SDSS training set. In our full WISE dataset, this cut eliminates about 600,000 sources, the vast majority of which are located within the Galactic plane and bulge, in Magellanic Clouds, and in M31; these are  regions of severe blending in WISE,  which is a further confirmation of the spurious nature of these \texttt{w1mag13<0} objects.

\begin{figure}
\begin{center}
    \includegraphics[width = 0.45\textwidth]{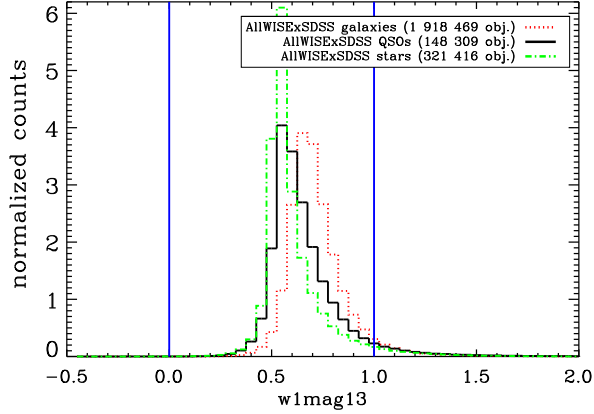}
  \end{center}
    \caption{Distribution of the $\mathtt{w1mag13}$ parameter (a proxy for flux concentration) for galaxies, stars, and quasars in the cross-match between AllWISE and SDSS spectroscopic sources. Vertical lines mark the position of the applied cuts  ($\mathtt{w1mag13}=0$ and $\mathtt{w1mag13}=1$) which remove blended sources from the training sample.}
 \label{fig:w13}
\end{figure}

\begin{figure}
\begin{center}
    \includegraphics[width = 0.4\textwidth]{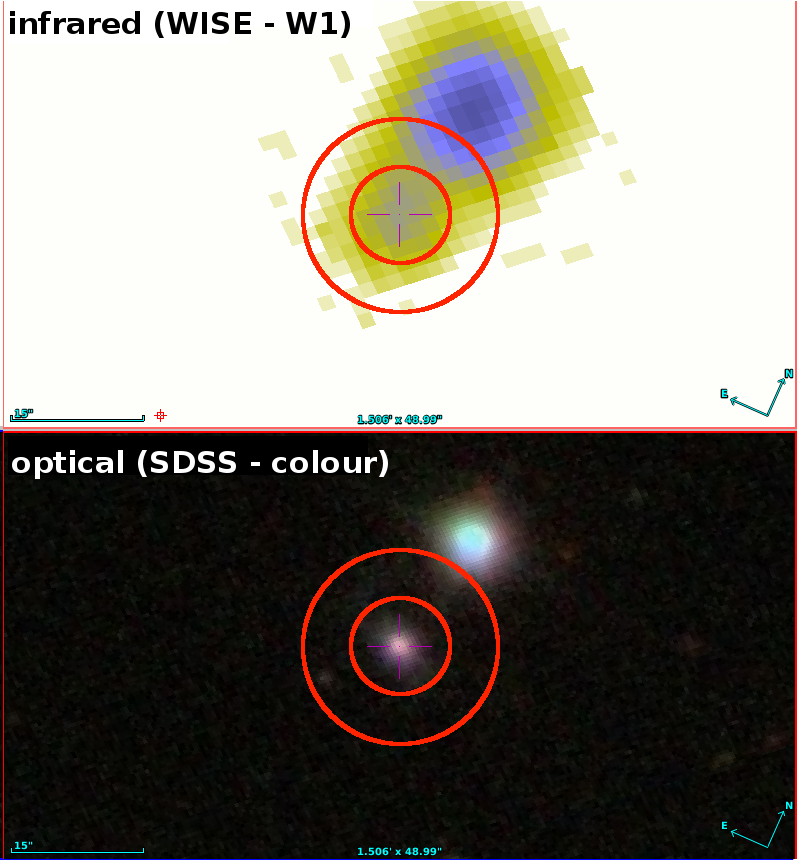}
  \end{center}
    \caption{Example of a quasar with a clean SDSS detection at $\alpha=192.00,\mbox{ }  \delta=10.17$ (lower panel, colour image constructed form $u$, $g$ and $r$ fiters) for which the WISE-derived concentration parameter ($\mathtt{w1mag13} \equiv \mathtt{w1mag\_1} - \mathtt{w1mag\_3}\;$) is  $1.13$ because of blending with a nearby star (upper panel, single-band image with $W1$ flux). Concentric circles mark the aperture in which the  $\mathtt{w1mag\_1}$ (5.5") and $\mathtt{w1mag\_3}$ (11") magnitudes were measured in WISE.}
 \label{w1mag13blend}
\end{figure}

Due to the much lower angular resolution of WISE  compared to SDSS ($6.1"$ in the $W1$ channel vs. $1.3"$ in the \textit{r} band), two objects appearing in close vicinity of one another can be well-separated in SDSS, but may be blended in WISE. This could then lead to high values of the \texttt{w1mag13} parameter, suggesting an extended source which in fact is a blend. Such objects would introduce biases during the anomaly detection process.
 The distribution of the \texttt{w1mag13} parameter in our training set is illustrated in Fig.~\ref{fig:w13}. It peaks at $\sim0.6$ for stars and quasars, and at $\sim 0.7$ for galaxies. Only a small fraction (2\%) of the training sources have $\mathtt{w1mag13}> 1$, and
we examined a representative sample of such objects by eye, starting from those with the most extreme values.
 We have found that the vast majority of them  are indeed blends, and this happens even if $\mathtt{w1mag13}\sim1$. An example  is shown in Fig.~\ref{w1mag13blend}. Two well-separated sources in SDSS (a quasar and a star) are blended in WISE and the $11"$ aperture centred on the object of interest gathers a large amount of flux from the second source. Such objects are not usable for the purposes of source separation and anomaly detection despite their usually excellent quality of SDSS measurements. Owing to these considerations, we will not be using objects with $\mathtt{w1mag13}>1$ for training; we then also have to remove such sources from the target catalogue. Such a cut removes a considerable number of AllWISE objects. However, for the purpose of the present analysis, these cuts are necessary as the training sample has to reflect the target sample in terms of parameter ranges. Otherwise target sources with parameter values differing  significantly from the input ranges would be automatically marked as anomalous.

\begin{table}
\centering
\caption{Summary of the training samples of SDSS objects used to train the OCSVM classifier.} 

\begin{tabular}{|c|c|c|}
\hline\hline
SDSS class & $N_{obj}$ before cuts & $N_{obj}$ after cuts \\
\hline
Galaxy &1~918~469& 1~827~211 \\
Star& 321~416&298~254\\
QSO & 148~309&141~471 \\ \hline\hline
\multicolumn{2}{c}{ }

\label{table:training}
\end{tabular}
\end{table}

After all the cuts discussed above, our training sample includes almost 2.3 million sources, of which 81\% are galaxies, 13\% are stars and 6\% are quasars (see Table~\ref{table:training}). 

\begin{figure}
\begin{center}

  \includegraphics[width = 0.5\textwidth]{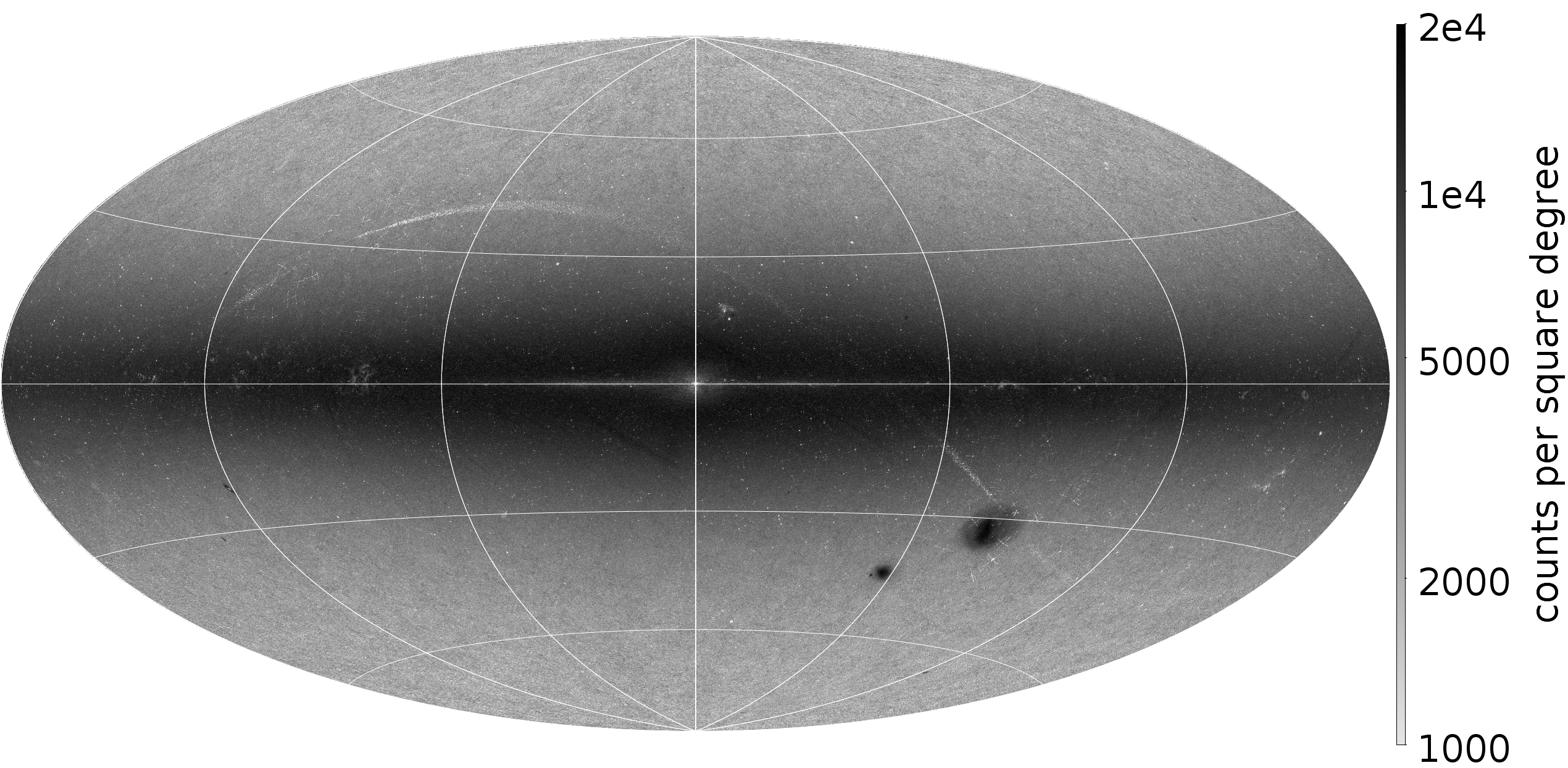}
     \end{center}
    \caption{Sky distribution of the 237 million AllWISE sources used for anomaly detection in this study. See text for details of sample preselection.}

 \label{fig:map_allwise}
\end{figure}

The final AllWISE catalogue that will be used for the novelty detection is composed of 237 million objects; see map in Fig.~\ref{fig:map_allwise}.  The most prominent features are our Galaxy and the Magellanic Clouds; however, there is  lower surface density in the Bulge, consistent with blending effects in areas of high projected density\footnote{\url{http://wise2.ipac.caltech.edu/docs/release/allsky/expsup/sec2_2.html}}. Also visible are stripes related to WISE instrumental issues\footnote{\url{http://wise2.ipac.caltech.edu/docs/release/allwise/expsup/sec2_2.html#w1sat}}.

\section{Novelty detection}
\label{sec:Novelty}

After the introduction of kernel methods \citep{vapnik,st}, pattern recognition schemes (ridge regression, e.g. \citealt{murphy12}; Fisher discriminant, e.g. \citealt{mika99}; principle component analysis, e.g. \citealt{scholkopf99}; spectral clustering, e.g. \citealt{langone15}; etc.) have gained in popularity in many branches of science where the amount of data being collected is increasing to the point where human processing is no longer practicable, which is the current situation in astronomy. Kernel methods explore linear and non-linear pair-wise similarity measures. Using non-linear kernels is equivalent to mapping data from the original input space onto a higher dimensional feature space where distinction between patterns can be easier.

This conventional pattern recognition  focuses on two or more classes. 
In a two-class problem we are dealing with a set of training examples $\mathbf{X}={(\mathbf{x_{i}},\omega_{i})|\mathbf{x_{i}}}\in \mathbb{R}^{D}\mbox{, }i=1...N $,  which contain $D$-dimensional vectors of $D$ characteristic properties (features or observables) for each of the $N$ examples (in astronomy: sources). Depending on the class the object belongs to, it is  then given a certain label $\omega=\{-1,1\}$. Next, out of the training dataset a function $h({\mathbf{x}})$ is constructed to estimate which label should be assigned to a new input vector \textbf{x'}: $\omega=h(\mathbf{x'}|X)$: 
\begin{equation}
h(\mathbf{x'|X}): \mathbb{R}^{D}\rightarrow[-1,1]\;.
\end{equation}
In the case of  classification schemes of more than two classes it is typical to decompose the problem into multiple binary problems. 
The final classification result combines partial outcomes of binary classifiers by a ranking method.
\noindent This conventional way, however, ignores any new/outlying data that do not belong to the considered classes. Without any freedom, the algorithm is forced to classify a source as one of the predefined classes, even if it does not fit to any presented category,  for example objects that do not occur in an optical-based training sample but are detected in the IR. 

To tackle the problem of novel data detection it is possible to modify the standard supervised classification scheme to \textit{one-class classification}. Here, the main class composed of normal/expected data points will be detected separately from all the other data points.
In the usual approach to novelty detection it is assumed that the normal class is well sampled, while the outlying class is undersampled. 
A model of normality $N(\theta)$ (not to be confused with normal
distribution), where $\theta$ is a free parameter of the model, is deduced and used to assign the novelty scores $n(\mathbf{x})$ to the previously unseen data \textbf{x}. In this sense, increasing scores can be understood as increasing deviation of the points from the normality model. 
We define the normality threshold as $z(\mathbf{x})=k$ in a way that an example \textbf{x} will be classified as normal if $z(\mathbf{x})\le k$ or as deviant in  the opposite case. Therefore $z(\mathbf{x})=k$ defines the {decision boundary}.
In this way the possibility of misclassification of the objects missing in the training sample is very low, as they will occur simply as deviations from {normal}.

To search for anomalies within the AllWISE data we have chosen to use a semi-supervised method belonging to knowledge-based algorithms (e.g. \citealt{scholkopf})  -- the support vector machines -- as these approaches focus on creating the decision boundary to contain the normal datapoints and are sensitive to outliers in both the training and test set. On the other hand, they do not depend on the distribution of the data within the training set;
 however, knowledge-based approaches to novelty detection have  one drawback: complexity associated with the computational time of kernel functions (see Sec.\ \ref{subsec:SVM}). Nevertheless, present-day technology coupled with parallelised computational capabilities significantly shortens the proper kernel choice for a given dataset and corresponding calculations.

There are also several other algorithms for novelty detection, such as reconstruction-based techniques such as neural networks \citep[e.g.][]{hawkins02,markou02} or subspace-based methods \citep[e.g.][]{jolliffe02,hoffmann07} that model the underlying data and reconstruct an error defined as a distance between the test vector and output of the system, which is then translated to a novelty score. 
However, even though reconstruction methods offer a flexible way to deal with high dimensionality of the data, they require a predefinition of parameters to define the structure of the model, which leads to two basic problems; the first  is the selection of the most effective training method to enable the integration of new units into the existing structure and the second  is the need to add a priori information about the saturation point (when no more new units can be added). 

Another large family of novelty detection techniques are distance-based approaches, which do not require any a priori knowledge about the data distribution, like nearest neighbour-based techniques \citep[e.g.][]{hautamaki04,angiulli09} or clustering-based techniques \citep[e.g.][]{cluseq,basu04}. However, they require a definition of distance metrics to establish similarity between data points, which becomes an increasingly persisting problem especially when dealing with high dimensionality of the parameter space (e.g. \citealt{kriegel09}) as distance measures in many dimensions lose ability to differentiate between normal and outlying data points. Moreover, these methods lack the flexibility of parameter tuning, making the methods unsuitable for full automation.

Owing to the above reasons, we chose to use support vector machines for our study; a detailed description of our approach   follows.

\subsection{Support vector machines}
\label{subsec:SVM}
Support vector machines is one of most commonly used conventional classification algorithms in astronomy. 
The idea of SVM is based on structural risk minimization (\citealt{vapnik74}). For many applications SVM have shown better performance and accuracy than other learning machines and have been used in many branches of astrophysics to solve classification problems and build catalogues \citep[e.g.][]{beaumont2011,fadely2012,malek13,solarz2015,KoSz15,heinis2016,marton2016,kurcz16,krakowski16}.
Support vector machines maps input points onto a high-dimensional  feature space and finds a hyperplane separating two or more classes with as large a margin as possible between points belonging to each category in this space.
Then the solution of the best hyperplane is composed of input points laying on the boundary called {support vectors} (SVs).

Here we outline the basis of the SVM theory in application to classification schemes.
Training of an SVM algorithm starts with having a set of observations with labels $(y_{1},\mathbf{x_{1}}),...,(y_{l},\mathbf{x_{l}})$, where $\mathbf{x_{i}}\in \mathbb{R}^{N}$ belongs to one of two classes and has a label $y_{i}\in \{-1,1 \}$ for $i=1,...l$. Each point should contain a vector of features,  characteristic values which describe it. Then the algorithm maps each vector from the input space $X$ onto a feature space $H$ using a non-linear function $\Phi:X\rightarrow H$. The desired separation plane $\mathbf{w}\cdot \mathbf{z}+b=0$ is defined by the pair ($\mathbf{w}$,b)  in such a way that each point $\mathbf{x_{i}}$ is separated according to a decision function 
\begin{equation}
f(\mathbf{x_{i}})=sgn(\mathbf{w}\cdot\mathbf{z_{i}}+b)
,\end{equation}
where $\mathbf{w}\in H$ and $b\in \mathbb{R}$.
 In principle, it is not the explicit knowledge of the mapping function $\Phi$ that is needed, but the dot product of the transformed points $\langle \Phi(x_{i}),\Phi(x_{j})\rangle$ \citep{cortes95}.
Therefore, instead of working with $\Phi$ it is possible to work with $K: X \times X \rightarrow \mathbb{R}$, where $K$ takes two points as input and returns a real value representing $\langle  \Phi(x_{i}),\Phi(x_{j}) \rangle$.
The only condition is that $\Phi$ exists  if and only if $K$ (called \textit{kernel}) is positive definite (satisfies Mercer's condition; \citealt{Mercer415}). 
Therefore, any function which meets this criterion can be a kernel function.
The most commonly featured kernel functions are linear, sigmoid, radial basis, and polynomial, which we describe in more detail in Section~\ref{sec:classscheme}.

\subsection{One-class SVM reformulation}
\begin{figure*}
\begin{center}
    \includegraphics[width = 0.8\textwidth]{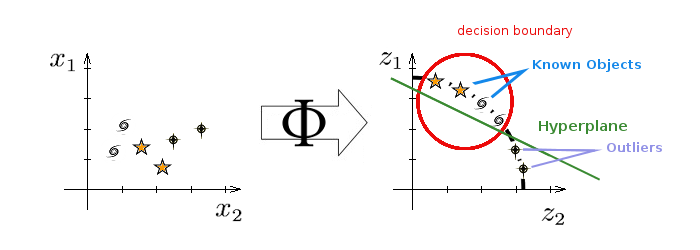}
  \end{center}
    \caption{Schematic representation of OCSVM using an example of the default radial basis kernel. The presented case of classification shows the tightest decision boundary which envelopes the known data (red circle) which can be treated as finding a separating hyperplane in the traditional SVM sense (green line). Unknown objects fall outside the sphere and are marked as outliers.}
 \label{schemat}
\end{figure*}

\citet{scholkopf} introduced an extension of the SVM methodology to pattern recognition  as an open set problem. Unlike the traditional SVM algorithm, which is designed to differentiate between classes contained within a given set, \textit{one-class SVM} (hereafter OCSVM) recognizes patterns in a much larger space of classes, unseen in training but which occur in testing. For that purpose, in the absence of a second class in the training data, the algorithm defines an `origin' by mapping feature vectors onto a feature space through an appropriate kernel function and then separates them by a hyperplane with a maximum margin with respect to the origin.
The resulting discriminant function is trained to assign positive values in the region surrounding the majority of the training points and negative elsewhere.
 Hyperplane parameters are derived by solving a quadratic programming problem
\begin{equation}
\mathrm{minimize}\left(\frac{1}{2}\mathbf{w}\cdot\mathbf{w}+\frac{1}{\nu l}\sum_{i=1}^{l} \xi_{i}-\rho\right)
\end{equation}
subject to 
\begin{equation}
\left(w\cdot \Phi(x_{i})\right)\ge \rho-\xi_{i};\, i=1,2,...,l;\, \xi_{i}\ge0\;,
\label{map}
\end{equation}
where \textit{w} and $\rho$ are parameters of the separation hyperplane, $\Phi$ is the mapping function of the input parameter space to a feature space, $\nu$ is the asymptotic fraction of outliers (anomalies) allowed, $l$ is the number of training points, and $\xi$ is a slack variable which penalizes misclassifications.
The decision function $f(x)=sgn(w\cdot\Phi(x)-\rho)$ determines point labels  (e.g. +1 for known instances and -1 for novel points). A schematic idea behind OCSVM is shown in Fig.~\ref{schemat}.

In this approach the parameter $\nu$ is interpreted as the asymptotic fraction of data labelled as outliers. The choice of the outlier fraction $\nu$ implies that the knowledge about the frequency of appearance of novel points is known a priori (e.g. \citealt{Manevitz07}). Otherwise, the value has to be tuned as a free parameter together with other unknowns. 
It is worth noting that domain-based approaches, such as this one, regulate the position of the novelty boundary using only those data with the closest proximity to it and that the properties of the distribution of data in the training set have no influence on this process \citep[e.g.][]{Tax99, TrungLe10, Le11, Liu11}. The only drawback of the presented method is the complexity associated with the choice and computation of the kernel functions. Moreover, the parameters controlling the size of the boundary area should be properly adjusted, increasing the computational time (e.g. \citealt{Tax2004}).   
In  this  work  we use the \textsf{R} \citep{rrr}\footnote{\url{http://www.R-project.org}} implementation of SVM included in the $\mathtt{e1071}$ package \citep{e1071}, which provides an interface to $\mathtt{libsvm}$. We use $\mathtt{doParallel}$\footnote{\url{https://cran.r-project.org/web/packages/doParallel/index.html}} and $\mathtt{caret}$\footnote{\url{https://cran.r-project.org/web/packages/caret/index.html}} packages to parallelize the computations

\subsection{Classification scheme}
\label{sec:classscheme}
To make a selection of outlying data it is crucial to create the best-suited classifier for a given dataset.
In our application, the classifier is trained on sources with spectroscopic measurements in the SDSS database treated as a single class, and present also in the AllWISE catalogue.
For this purpose  we include all sources from the AllWISE$\times$SDSS cross-match in the training sample. Unlike in the case of classical SVM, the imbalance of the training set has no influence on the OCSVM training, as we create only one known class. The quantity and ratios of specific classes are not an issue here. For that reason OCSVM can also be treated as an alternative approach to dealing with imbalanced datasets which offers no information loss during training (e.g. \citealt{imbalance}).
With the training sample selected, the algorithm has to be trained to recognise the {normal} patterns, which in the case of OCSVM means  finding the best-suited volume encompassing the training points, which  will later be used as a decision boundary between what the algorithm finds as normal and deviating patterns.
 This procedure involves searching for the most appropriate  kernel function, which governs the topology of the surface enclosing  the training sample. 
To ensure the best performance of the algorithm it is necessary to choose an appropriate kernel function for the given training set and to find its meta-parameters to train the novelty detector which will best suit the input data. As no two datasets are the same, it is natural that there is no universal kernel function optimal for each classification problem. This makes testing several functions a vital step in any kernel-based machine learning process \citep{Sangeetha2010}. 

In this study we test four basic shapes of kernel functions in the application of the novelty detection to the AllWISE dataset
\begin{enumerate}[i)]
\item linear kernel: $u^{T}v$, 
\item sigmoid kernel: $\tanh(\gamma u^{T}v + C),$
\item radial basis kernel: $\exp\left(-\gamma||u-v||^{2}\right)$,
\item polynomial kernel: $(\gamma u^{T}v + C)^{d},$
\end{enumerate}
where $u$ and $v$ are vectors in the input space, $||\cdot||^{2}$ is the squared Euclidean distance between the two feature vectors, $\gamma$ is a scaling parameter, $d$ is the degree of the polynomial function, and $C$ is a constant. These meta-parameters need to be tuned for each given dataset; an exception is the linear kernel which does not have any free parameters.

Taking into  account all the above points, the following  steps need to be applied for each considered kernel function in order to determine the best-fitting
topology of the separation hypersurface for any dataset:
\begin{enumerate}

        \item \textbf{Division of the training set}:\\
 The full training set is divided into two subsets, where one is used for the actual training and the other is used as a validation subset to verify the accuracy of the created hypersurface against sources with known class which were not used by the algorithm to find the model. We create the training set out of a random 99\% 
 of known objects;  the validation set contains the remaining 1\% of known sources. The percentage left for validation is small but sufficient to verify whether the classifier works well on previously unseen data.
        \item \textbf{Wide grid search}:\\
Training a classifier means  finding the best set of kernel function meta-parameters. They are determined by searching through a loosely spaced grid of meta-parameter values describing each kernel (e.g. $d$, $\gamma$, and $C$ in the case of the polynomial kernel) and the $\nu$ parameter specifying the expected outlier fraction.
        \item \textbf{Estimation of training accuracy:}\\
For each tested combination of meta-parameters we count how many times an object with known nature was correctly classified by the OCSVM (true positive; TP) and how many times a known object was classified as an outlier (false negative; FN). Based on these counts {accuracy} is calculated as $acc_{train}=TP/(TP+FN)$.
 Moreover, we count the number of SVs used to  find the decision boundary. The fewer the points  treated as SVs, the better: when the data is well-structured only  a small number of SVs are used;  all the remaining training points will not be used in the calculation of the final boundary. High numbers of SVs mean that the topology of the surface is complex, and that the data cannot be easily contained within the boundary.

        \item \textbf{Estimation of validation accuracy:}\\
        The algorithm, trained in the previous step, with its best-suited meta-parameters of its kernel, is applied to classify the validation set. Thanks to the knowledge of the true labels of this set it is possible to verify how well the trained algorithm is working on previously unseen data by calculating $acc_{valid}=TP/(TP+FN)$ and therefore to estimate how well it will work on truly unknown data. As above,   the number of SVs is also taken into account here.
        \item  \textbf{Fine-tuning of grid search}:\\
        We tighten the grid search around the best values from the wide grid to fine-tune the meta-parameter choice for the best performance (by repetitive measurements of $acc_{train}$, $acc_{valid}$). 

\end{enumerate}

The search for the free parameters of each kernel is done within reasonable expected ranges.
To find  these ranges we follow the scheme of \citet{chapelle05}, where it is first necessary to fix initial values for each set of parameters which will provide the most reliable orders of magnitude. In the case of the one-class problem we use the median
of pairwise distances of all training points as the default for $\gamma$.
 The default for $\nu$ is taken as the inverse of the empirical variance $s^{2}$ in the feature space calculated as $s^{2}=\frac{1}{n}\Sigma_{i}K_{ii}-\frac{1}{n^{2}}\Sigma_{i,j}K_{i,j}$ from an $n \times n$ kernel matrix $K$.
For the degree of the polynomial we consider only two possible values, 2 and 3, as higher degrees would create a boundary whose topology is  too complex, which would result in overfitting  the model. 
Then we use multiples ($10^{k}$ for $k\in\{-3,...,3\}$) of the default values as the grid search range. 

\begin{table}
\centering
\caption{ Input parameter ranges for the grid search for the kernels tested in Sec.\ \ref{sec:Results} (see text for details). }
\label{ajaj}
\begin{tabular}{|c||c|c|c|c|}
\hline\hline
Kernel&$\nu$&$\gamma$&$C$&$d$\\
\hline
linear&0.0001-0.69&-&-&-\\
radial&0.0001-0.69&0.001--10000&-&-\\
sigmoidal&0.0001-0.69&0.001--10000&0--4&-\\
polynomial&0.0001-0.69&0.001--10000&0-4&2-3 \\ \hline\hline
\multicolumn{5}{c}{ }
\label{kerneleoneclass_iter}
\end{tabular}
\end{table}

After pinpointing the best set of parameters for each kernel function we perform fine-tuning of the grid search, where we search the grid around those best parameters with a much smaller step (multiples of $2^{k}$).
 However, we find that the maxima of the performance around the most optimal parameters found on the sparser grid are very broad and the fine-tuning of the grid does not improve the performance of the classifier.
Ranges of searched parameter values  for each kernel are summarized in Table~\ref{kerneleoneclass_iter}.

Upon finalization of the training and verifying that the classifier performs on a satisfactory level, it is applied  to the AllWISE target data to search for true outlying objects.

\section{Results of OCSVM application to AllWISE data }
\label{sec:Results}
In this section we present the results of applying the
OCSVM algorithm to the AllWISE catalogue. 
Following the discussion presented in the previous section, we started by determining the most appropriate kernel for this training set. We found that in this case the preferred kernel function is radial-based with optimal parameters $\gamma=0.1$ and $\nu=0.001$ as it provides the highest training and validation accuracies and is characterized by the smallest number of SVs. 
The optimal parameters and training/validation performance of all four tested kernels are summarised in Table~\ref{kerneleoneclassall}. 

\begin{table}
\caption{Performance of the kernel functions tested in the training of the OCSVM algorithm.} 
\begin{center}
\begin{tabular}{|c||c|c|c|c|c|c|c|}
\hline\hline
Kernel&\multicolumn{4}{c|}{best parameters}&$N_{SV}$&\textit{$acc_{train}$}&\textit{$acc_{valid}$}\\\hline
&$\nu$&$\gamma$&$C$&$d$&&&\\\hline
linear&0.1&-&-&-&892&27.02\%&31.33\%\\
sigmoid&0.001&1&0&-&125&99.99\%&98.80\%\\ 
radial&0.001&0.1&-&-&48&99.99\%&99.98\%\\ 
poly&0.005&100&0&3&53&97.87\%&96.66\% \\ \hline\hline
\multicolumn{8}{c}{ }
\label{kerneleoneclassall}
\end{tabular}
\end{center}
\end{table}

\begin{figure}
  \centering
\includegraphics[width=\linewidth]{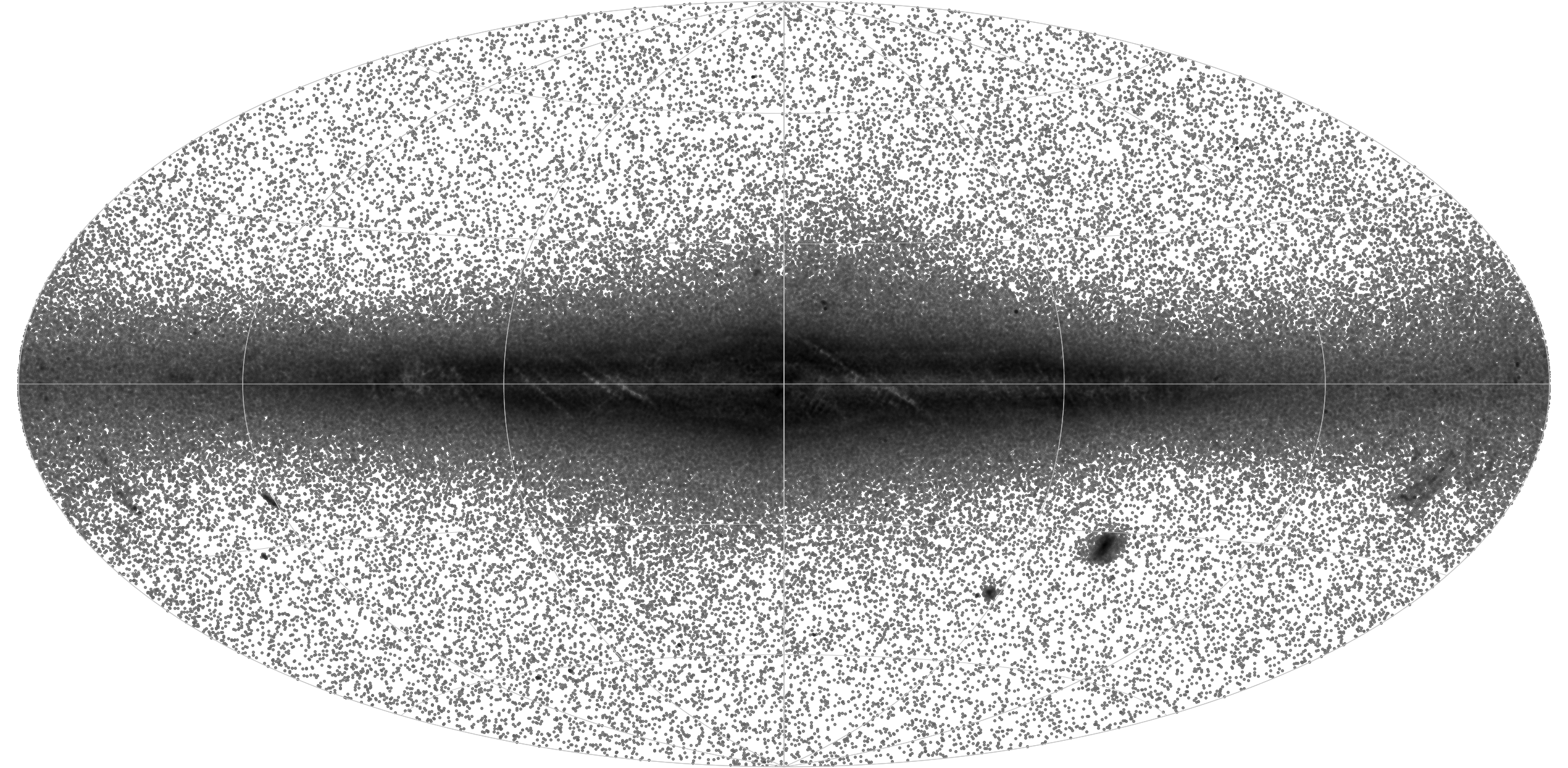}
  \caption{Sky positions of all the objects classified as {unknown} in the application of the OCSVM anomaly detector to the AllWISE catalogue (642~353 sources), shown in Galactic coordinates.}
\label{allskypos}
\end{figure}

Having determined the proper kernel, we trained the OCSVM anomaly detector on the AllWISE$\times$SDSS training set, and subsequently applied it to AllWISE data preselected as described in Sec.~\ref{sec:Data}. As a result, we obtained a sample of 642,353 sources classified as {unknown} by the algorithm. We show their sky distribution in Fig.~\ref{allskypos}; as is obvious from the plot, the vast majority of the sources is located within the Galactic Plane and Bulge (90\% are within $|b_\mathrm{Gal}| < 15^\circ$) and in other confusion areas: Magellanic Clouds, Galactic dust clouds, and even M31 and M33. This  is to be expected as the $6"$ spatial resolution of the WISE satellite leads to severe blending in areas of high projected density, which in turn results in anomalous (spurious) photometric properties of these blended objects. However, as discussed below, except for such artefacts, our anomaly detector also flagged  a considerable number of genuine sources of astrophysical interest.

\begin{figure}
  \centering
\includegraphics[width=.95\linewidth]{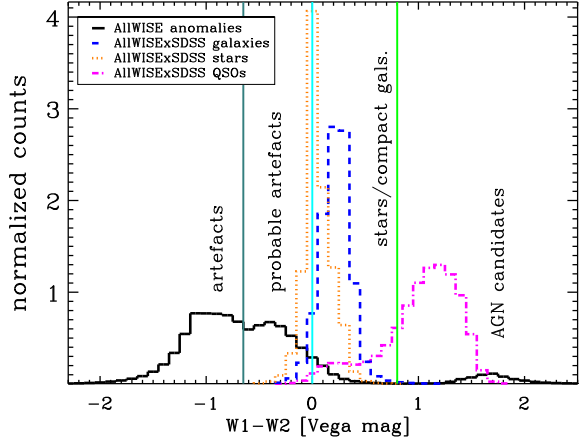}
  \caption{Distributions of the $W1-W2$ colour for AllWISE anomalies found in this work (solid black) compared with  {known} sources from AllWISE$\times$SDSS used to train the OCSVM algorithm. Galaxies (1~827~241 objects) are marked by blue dashed lines; stars (298~269 objects) by orange dotted lines; and quasars (141~494 objects) by magenta dot-dashed lines. Vertical lines mark the colour cuts applied to the OCSVM anomaly sample dividing it into four possible subgroups.}
\label{fig:w1w2_anomalies}
\end{figure}

To gain insight into the nature of these anomalies, we started by looking at their WISE colours. 
It is important to note that the OCSVM algorithm itself does not provide any means of discriminating various populations among the outliers. Relying on the colours to identify the groups in the resultant anomalies is the most basic approach. It is possible to refine this task by employing clustering algorithms (e.g. \citealt{clust}) which could find different classes within the outlier group. For the bright sources the problem can  also be approached by using passband images directly (e.g. \citealt{hoyle15}), but the speed of data processing would significantly decrease. However, in this work we restrict ourselves to a first look at the anomalies, and we mostly use a single WISE colour, $W1-W2$, for that.
 In Fig.~\ref{fig:w1w2_anomalies} we present their $W1-W2$ distribution and compare it to the training set, divided according to SDSS source classes (stars, galaxies, quasars). We observe multi-modal behaviour of this colour for the detected anomalies, with three peaks at $W1-W2\sim -1$, $\sim-0.5$, and  $\sim1.7$. The peaks are separated by minima at $W1-W2 = -0.65$ and at $\sim0.8$.  It is interesting to note that the latter is the same as the WISE active galactic nuclei (AGN) separation criterion first proposed by \cite{stern12}; we discuss these red sources in more detail in Sec.~\ref{sec:QSOs}.
The total number of sources contained in each considered group is shown in Table~\ref{table:anomalies}.

\begin{figure*}
\centering
\includegraphics[width=\linewidth]{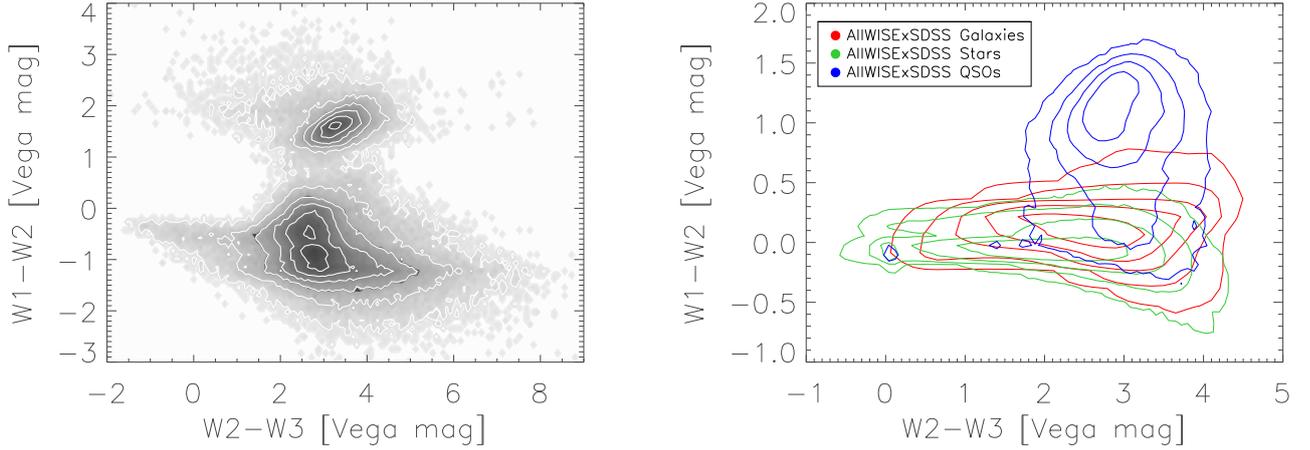}
\centering
\caption{Left panel: WISE colour-colour ($W1-W2$ vs. $W2-W3$) diagram for the sources identified by OCSVM as outliers in AllWISE. The plot shows only sources detected in the $W3$ band (188~496 objects comprising 38\% of all the anomalies). The grey scale marks the density of displayed points in linear bins. Right panel: $W1-W2$ vs. $W2-W3$ diagram for AllWISE$\times$SDSS sources used for OCSVM training. We note the different ranges of the axes in the two panels.}
\label{fig:CC_anomalies}
\end{figure*}

This division into roughly three groups is also confirmed  in the CC diagram where the $W2-W3$ WISE colour is used as the second dimension (Fig.~\ref{fig:CC_anomalies}). To construct this diagram we used only those sources which had positive signal-to-noise ratio in the $W3$ band, which is 38\% of the full anomaly sample. In addition, for objects with $0<\mathtt{w3snr}<2$, which have only $W3$ upper limits in the WISE database, we applied the correction of $+0.75$ mag as discussed in the Appendix of \cite{krakowski16}. This CC diagram gives indications of the nature of the detected anomalies: according to fig. 26 in \cite{jarrett11} or fig. 5 in \cite{cluver14}, stars concentrate around $W1-W2 \simeq W2-W3 \simeq 0$, elliptical galaxies have $W1-W2 \gtrsim 0$, and $0.5 < W2-W3 < 1.5$ while spirals span $0 < W1-W2 < 0.5$ and $1 < W2-W3 < 4.5$; quasars are much redder in $W1-W2$ than most of inactive galaxies while their $W2-W3$ is similar to that of some spirals. There are also some more specific sources located on that diagram, such as (U)LIRGs or brown dwarfs, but here we will restrict ourselves to the basic three classes (stars, galaxies, and quasars) in our basic division of the anomalies.
Compared with  the theoretical $W1-W2$ vs. $W2-W3$ diagram, we see that the upper cloud of our anomalies is located roughly at the (obscured) AGN locus, while the two lower ones do not seem to be consistent with any normal sources in this plane.

 Below we discuss in more detail the possible nature of these three groups of sources. We reiterate that as most of the detected anomalies do not have measurements in the $W3$ band, the distinction will be made only based on their $W1-W2$ colour.

\begin{table}
\centering
\caption{Number of sources identified as anomalies by the OCSVM algorithm, further divided  according to the $W1-W2$ colour cuts.} 

\begin{tabular}{|c|c|}
\hline\hline
colour cut & $N_{obj}$ \\
\hline
$W1-W2 < 0$ & 575~598\\
$0\le W1-W2<0.8$ & 26~990 \\
$W1-W2 \ge 0.8$ & 39~940\\ \hline\hline
\multicolumn{2}{c}{ }

\label{table:anomalies}
\end{tabular}
\end{table}

\subsection{Anomalies with extremely low $W1-W2$ colour: photometric artefacts}
\label{sec:blends}

\begin{figure}
  \centering
\includegraphics[width=\linewidth]{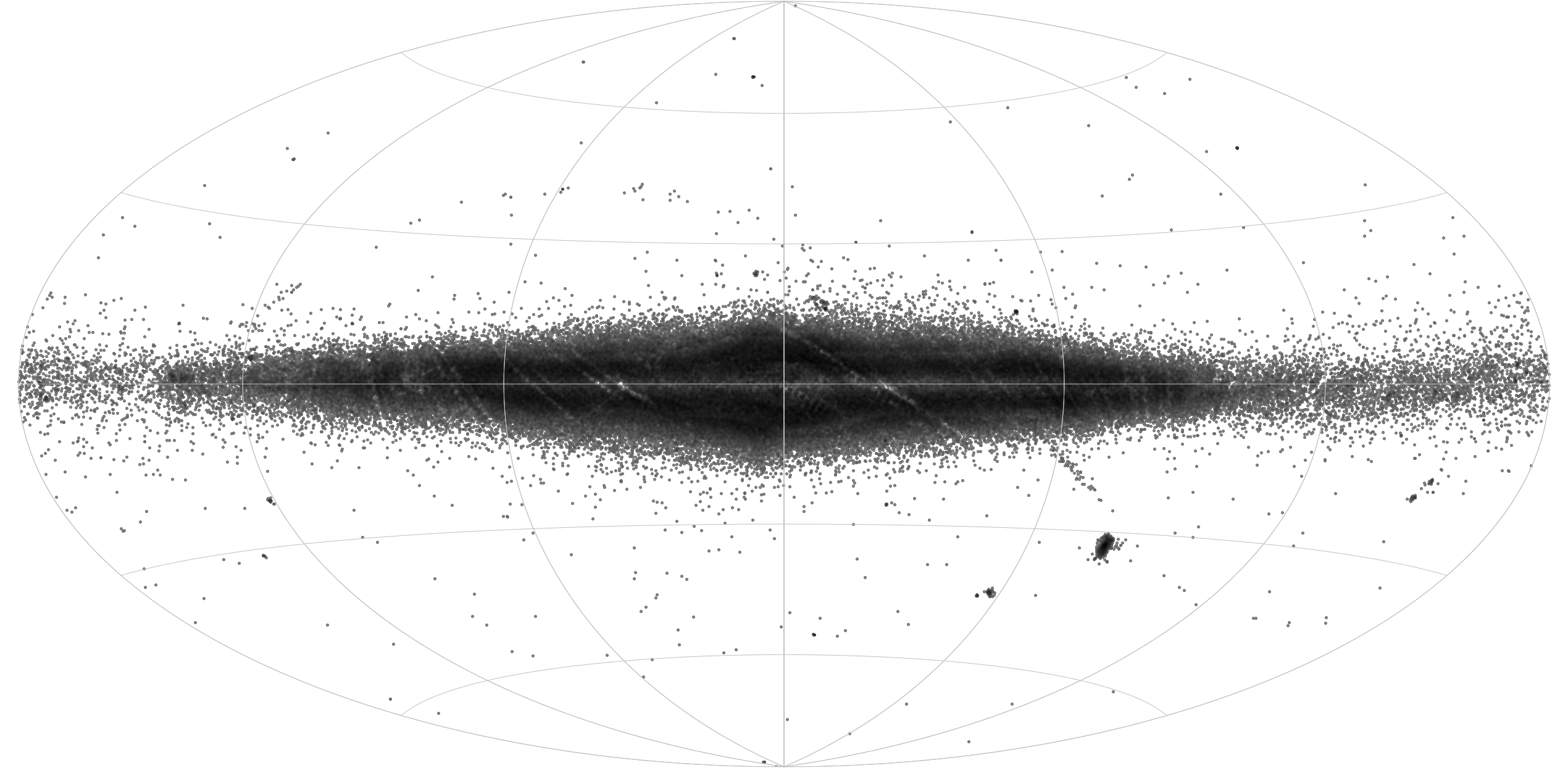}
  \caption{Sky distribution of AllWISE anomalous sources with $W1-W2 < -0.65$ (354~301 objects) shown in Galactic coordinates.}
\label{fig:sky_blends}
\end{figure}

We begin our detailed investigation into the nature of the detected anomalies by looking at those with extremely low $W1-W2<-0.65$. This is the majority (55\%) of the outliers found by OCSVM. Their sky distribution (Fig.~\ref{fig:sky_blends}), and the fact that they have such an extremely blue and unphysical $W1-W2$ colour, is a clear indication that these are sources with spurious photometry due to blending. In what follows we will not deal with these objects further. We note, however, that their identification by our anomaly detector is  evidence of its successful operation. 

Even after removing the $W1-W2<-0.65$ artefacts, there is still a significant fraction  of sources with a very negative and most likely non-physical $W1-W2$ colour (cf.\ Fig.~\ref{fig:w1w2_anomalies}). Almost all such sources are confined to the Galactic plane and bulge; to inspect whether they have indeed spurious photometry, we checked their properties in GLIMPSE \citep{glimpse},\footnote{We used GLIMPSEII 2.1 Data Release, \url{http://www.astro.wisc.edu/glimpse/glimpse2_dataprod_v2.1.pdf}.}  a Spitzer survey covering the inner Galactic Plane and Bulge within $|b_\mathrm{Gal}|<5^\circ$ \& $|l_\mathrm{Gal}|<65^\circ$.
As filter coverage of Spitzer IRAC $I1$ (centred at 3.6~$\mu$m) and IRAC $I2$ (centred at 4.5~$\mu$m) are very comparable to WISE $W1$ and $W2$, respectively \citep{jarrett11}, this is an adequate test to confront corresponding flux measurements from the GLIMPSE catalogue with those of the OCSVM AllWISE anomalies.

\begin{figure}
  \centering
\includegraphics[width=\linewidth]{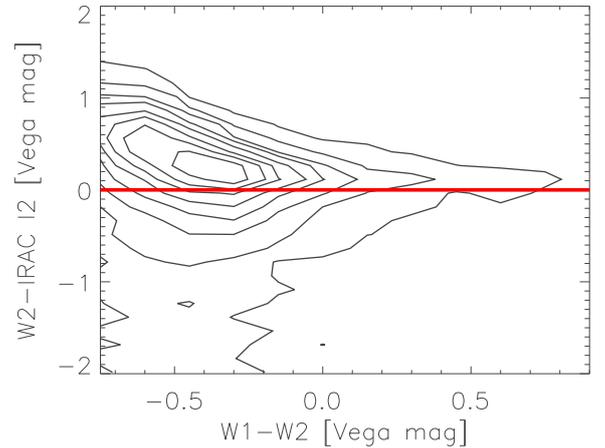}
  \caption{ Difference between Spitzer IRAC $I2$ and WISE $W2$ magnitudes for the anomalous sources with $W1-W2<-0.65$ (354~301 sources) as a function of the WISE $W1-W2$ colour. The $I2$ measurements were taken from the GLIMPSE survey of the Galactic Plane. Contours mark the density of the displayed points in linear bins. } 
\label{fig:W2bias}
\end{figure}

Our sample of outliers with $0> W1-W2>-0.65$ has almost 17~000 counterparts in GLIMPSE within a $3"$ matching radius. By comparing the IRAC $I1$ \& $I2$ vs. WISE $W1$ \& $W2$ measurements, we found that  for the shorter-wavelength channels the IRAC and WISE magnitudes match very well, while in the case of $I2$ vs. $W2$ comparison there is a clear discrepancy: WISE measurements in this band significantly underestimate the fluxes with respect to IRAC. What is more, this bias increases with decreasing $W1-W2$ colour (Fig.~\ref{fig:W2bias}). 
Sources with $0>W1-W2>-0.65$, however, could  hide a fraction of real sources of astronomical interest as objects with moderately negative $W1-W2$ colours have been reported  \citep[e.g.][]{banerji13, jarrett17}. Current parameter space does not allow for a proper distinction between the real and spurious sources within this anomaly group; only a more in-depth analysis with clustering algorithms could reveal more insight on that matter. As this task is beyond the scope of this work,  in the present approach we restrict ourselves to treat all anomalies with $W1-W2<0$ either as having spurious $W2$ photometry or as problematic, and we also remove them from further examinations. 
This cut only affects   confusion areas (Galaxy, Magellanic Clouds) and significantly purifies the anomaly dataset.

\subsection{Anomalies with $W1-W2>0.8$: AGN/quasar candidates}
\label{sec:QSOs}
We now turn our attention to the group of sources with $W1-W2>0.8$, which clearly stand out in the CC diagram (Fig.~\ref{fig:CC_anomalies}); there are almost 40,000 such anomalies in our sample. As already mentioned, their location in this diagram is consistent with them being AGN/QSO, i.e.\ high-redshift extragalactic sources. There are several lines of evidence supporting this hypothesis.

\begin{figure}
  \centering
\includegraphics[width=\linewidth]{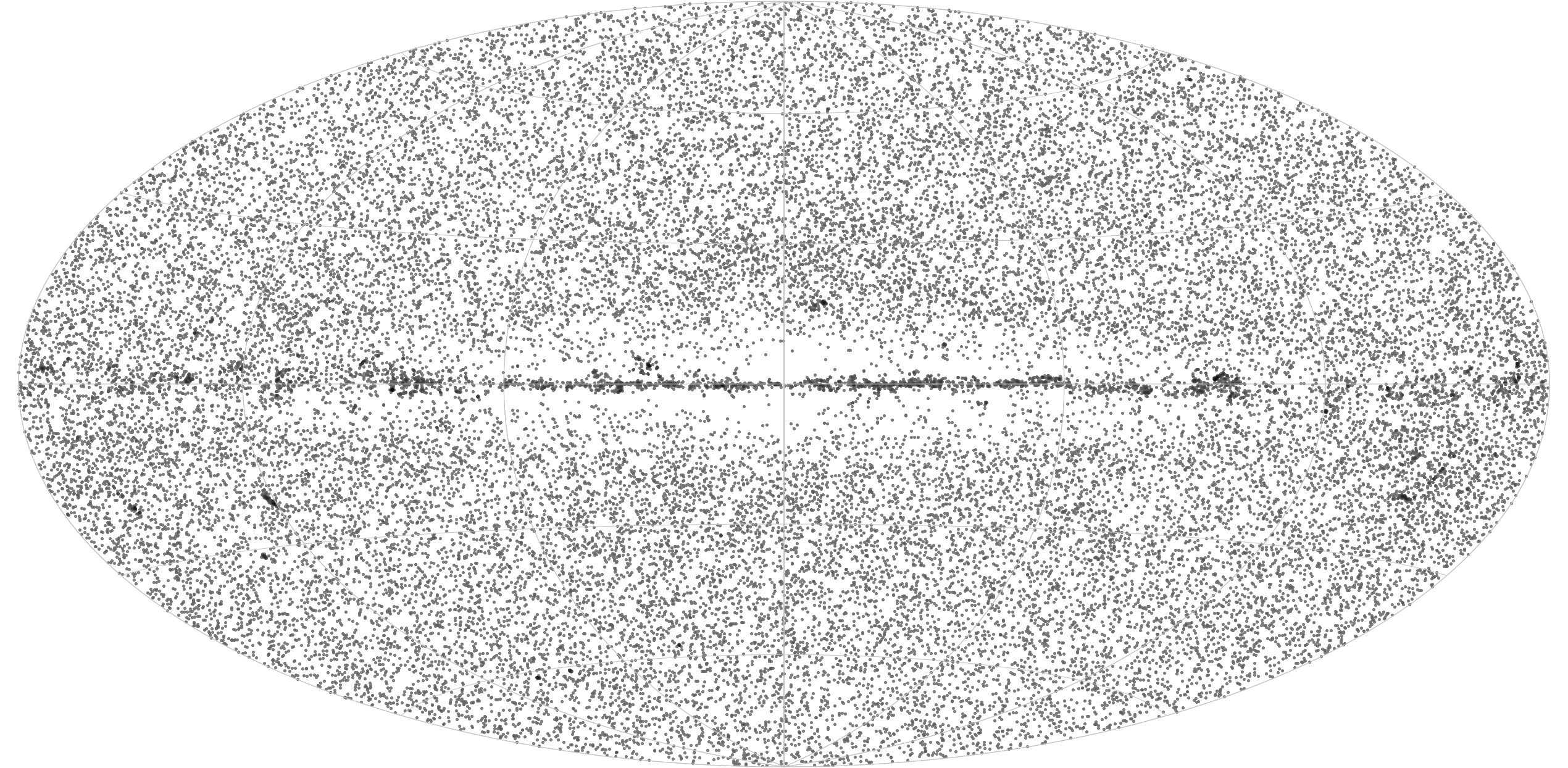}
  \caption{Sky distribution of AllWISE anomalous sources with $W1-W2 \ge 0.8$ (39~940 objects).}
\label{fig:sky_topcloud}
\end{figure}

\begin{figure}
  \centering
\includegraphics[width=\linewidth]{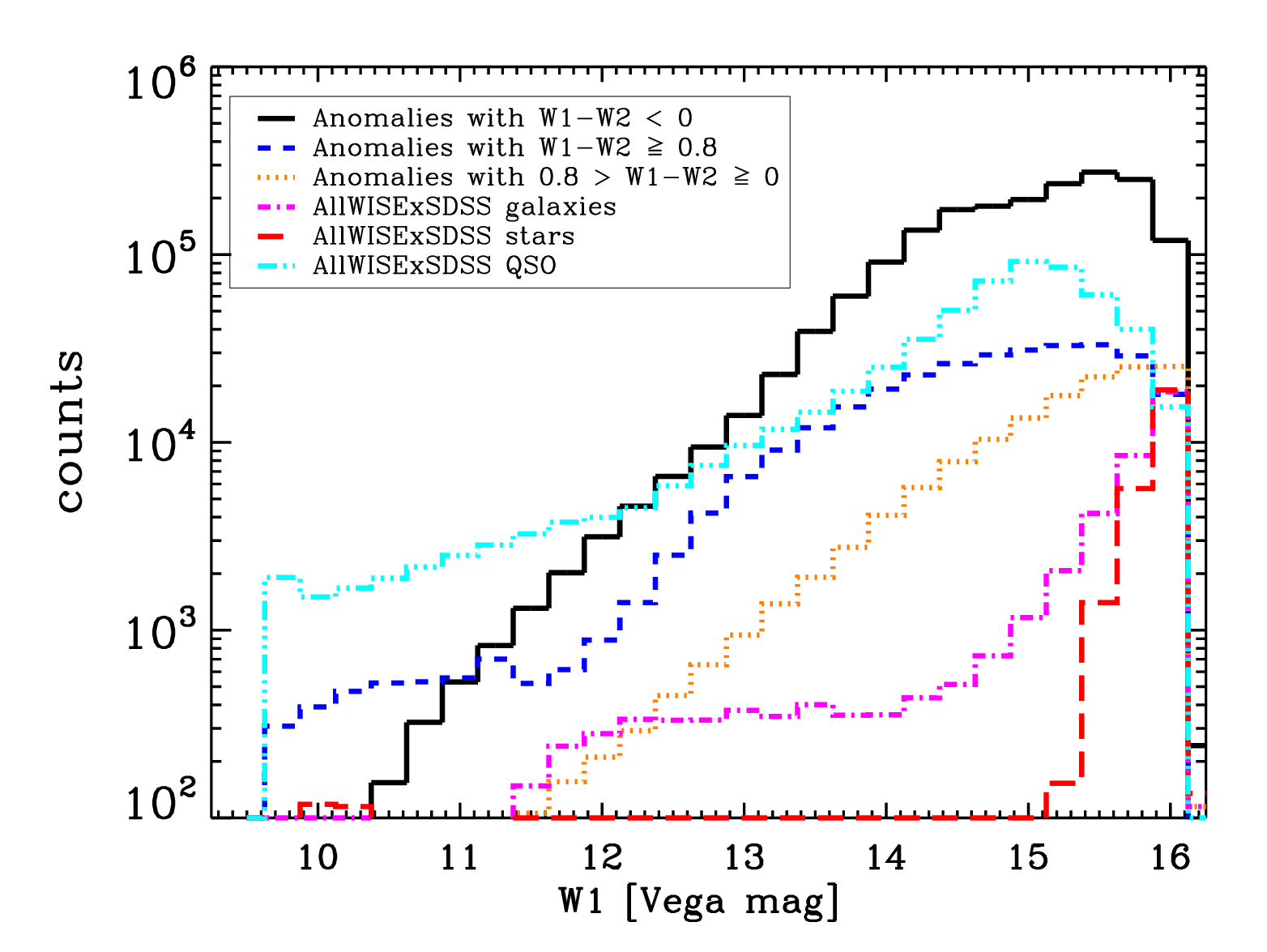}
  \caption{$W1$ magnitude distributions for the three main types of anomalies identified by OCSVM in AllWISE: with $W1-W2<0$, solid black (575~423 sources); with $0\le W1-W2<0.8$, orange dotted (26~990 sources); with $W1-W2\ge 0.8$, blue dashed (39~940 sources), compared with  the {known} sources from AllWISE$\times$SDSS used to train the OCSVM algorithm: galaxies, magenta dot-dashed (1~827~241 sources); stars, red dashed (298~269 sources); and quasars, cyan triple-dot-dashed (141~494 sources).}
    \label{fig:W1_all_anomalies}
\end{figure}

First of all, these sources are very uniformly distributed over the entire sky (Fig.~\ref{fig:sky_topcloud}), and are preferentially located {outside} the Galactic Bulge, except for some at the Galactic equator (5,200 within $|b_\mathrm{Gal}|<3^\circ$)  which must be artefacts of WISE blending in a similar way as the very low $W1-W2$ sources of Sec.~\ref{sec:blends}. Secondly, these anomalies are mostly faint: their $W1$ counts peak at the limit of the catalogue used here, $W1=16$ (see Fig.~\ref{fig:W1_all_anomalies}). Furthermore, almost all of them (over 95\%) have WISE detections in the $W3$ channel ($\mathtt{w3snr}>0$). We reiterate that this channel was not used in the source preselection procedure; thus, as its sensitivity in WISE is much lower than that of the two shorter-wavelength bandpasses, this indicates that these sources are intrinsically bright at (observed) 12 $\mu$m. 
As we can quite safely exclude the situation in which stellar light would be redshifted to this channel (one would need $z>2$, which for $W1<16$ would mean intrinsic brightness of $\sim-30$ mag or brighter at rest-frame $\lambda\sim1\;\mu$m), this points to emission from dust as 12 $\mu$m observations are sensitive to warm dust radiation \citep[e.g.][]{sauvage05} and polycyclic aromatic hydrocarbon  emission lines \citep[PAH, e.g.][]{pah} at redshifts lower than 2.
On the other hand, a cross-match with the all-sky 2MASS data \citep{2mass06} gave only 1500 sources, most of which in its \textit{Point} Source Catalogue (PSC) and just a handful (45) are extended (in 2MASS XSC, \citealt{jarrett00}). This shows that most of our QSO candidates are not in the local volume, as 2MASS provides a very complete census of the local Universe \citep{2MPZ,rahman16}.


Further insight into the nature of these objects is gained by checking for their presence and properties in the SDSS {photometric} catalogue\footnote{Avaliable at \url{http://skyserver.sdss.org/dr13}}.  For training we used only sources with SDSS spectroscopy, while the general photometric dataset from Sloan is obviously much larger and more complete  at a price of much more limited information on the real nature of the detected sources. By cross-matching with SDSS photometric, we found about 7000 of our  AGN candidates to have counterparts there within a matching radius of $1"$. About 3000 of them are also present in the SDSS DR12 photometric redshift catalogue of \cite{beck16},  which contains SDSS-resolved galaxies up to $z=1$ -- and these matched objects have mean $\langle z \rangle \sim 0.5$. By extrapolating to the full extragalactic sky, this exercise suggests that about 40\% of these anomalies would have no optical counterpart in an SDSS-depth all-sky catalogue if one existed. On the other hand, about 25\% of these  AGN candidates seem to be residing in optically resolved galaxies at $z<1$.

\begin{figure}
  \centering
\includegraphics[width=0.9\linewidth]{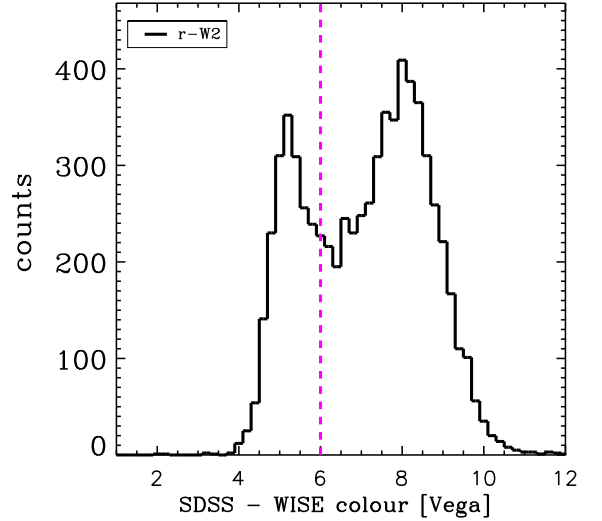}
  \caption{Histogram of the $r-W2$ colour for the AGN candidates identified by OCSVM which have a counterpart in the SDSS photometric catalogue (7~053 sources). The vertical line marks the division  between obscured and unobscured AGNs according to \cite{donoso14}. The SDSS $r$-band magnitudes were converted to the Vega system following \cite{blanton07}.} 
\label{fig:rW}
\end{figure}

\begin{figure}
  \centering
\includegraphics[width=.95\linewidth]{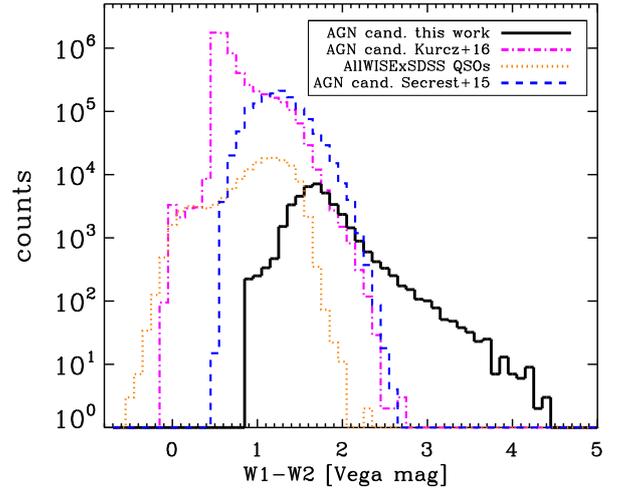}
  \caption{Histograms of the $W1-W2$ colour for quasars and quasar candidates from the following datasets: AllWISE paired up with SDSS DR13 spectroscopic \citep[dotted orange; 141~494 sources;][]{sdssdr13}; AllWISE OCSVM AGN candidates (black solid line; 39~940 sources; this paper); AllWISE SVM AGN selection \citep[magenta dot-dashed; 4~443~962 sources;][]{kurcz16}; and AllWISEAGN \citep[blue dashed; 1~354~775 sources;][]{agnallwise}. }
\label{fig:qsotrainingw1w2}
\end{figure}

According to studies of AGNs identified in both WISE and SDSS   
\citep[e.g.][]{yan13,donoso14}, the combined optical-MIR $r-W2$ colour can be used as a diagnostic to differentiate between unobscured/type-1 and dust-obscured/type-2 AGN/QSO candidates, the division being $r-W2\sim6$ (both Vega). We have thus checked the behaviour of this colour in our sample of anomalies with $W1-W2>0.8$ also present  in SDSS. Indeed, we observe bimodality in the $r-W2$ colour with the division roughly at $r-W2\sim6$ (Fig.~\ref{fig:rW}); a similar bimodality is also present in the distribution of the $r-W1$ colour.
This suggests that the OCSVM selects both types of AGN populations, although we emphasise that their $W1-W2$ colour is in most cases much redder than for the AllWISE$\times$SDSS spectroscopic QSOs, used as part of the training (Fig.~\ref{fig:qsotrainingw1w2}).

In the next step we compared our findings with other WISE-based QSO candidate selections. We will limit ourselves to those works which presented all-sky WISE data in this context, namely \cite{agnallwise} and \cite{kurcz16}.
 The OCSVM-selected QSO candidates have much redder IR colours than the QSO candidates of \citet{kurcz16}, but also redder than the quasars in the SDSS-based training sample used both here and in that paper (cf. Fig.~\ref{fig:qsotrainingw1w2}). In addition, they do not show the anomalous sky distribution present in \cite{kurcz16} (non-uniform distribution on the sky with somewhat larger surface density close to the ecliptic than at the ecliptic poles). 
There are over 22~000 common sources between the OCSVM-QSO sample and the \cite{kurcz16} AllWISE QSO candidate one, and the common sources have $W1-W2 < 1.8$. Beyond those values, however, OCSVM selects much redder  QSO candidates, missed by the classical approach of source classification: in \cite{kurcz16} the very high $W1-W2$ objects were assigned to random classes and had roughly equal probabilities of belonging to any of them. 

Finally, we compared the results of the OCSVM QSO selection with the publicly available 
AllWISEAGN catalogue \citep{agnallwise} containing over 1.4 million AGN candidates  extracted from AllWISE following the formulae of \citet{mateos12}.  In addition to another method of AGN/QSO selection -- colour-based vs. automated -- the AllWISEAGN sample also uses  different preselection criteria to ours. Namely, \cite{agnallwise} required all their sources to have $S/N\geq5$ in all the first three WISE channels, while we do not use the $W3$ band for selection or classification. We note that due to AllWISE observational limitations, the 12 $\mu$m $S/N$ requirement of \cite{agnallwise} leads to very non-uniform sky coverage of their AGN candidates, varying over an order magnitude in surface density on different patches of the sky (Figs.~1 \& 2 therein); no such issues are evident in our sample except for the Galactic equator area. On the other hand, our  QSO candidate sample is much shallower than the AllWISEAGN one because of our requirements of $W1<16$ and $W2<16$, while no such cuts were applied in \cite{agnallwise}; the latter sample thus  reaches formally to the full depth of AllWISE (modulo the additional $W3$ preselection), which is $W1\sim17$ on most of the sky, and over $W1=18$ by the ecliptic poles \citep{jarrett11}.
 There are over 25~000 common sources between our QSO candidate dataset and AllWISEAGN, which means that almost 30\%  of our sample outside the Galaxy was not identified by \cite{agnallwise} as AGN candidates. Taking into account that our dataset is one magnitude shallower than AllWISEAGN, we expect that applying our method at the full depth of AllWISE will bring of the order of 100~000
 more QSO candidates not contained in the \cite{agnallwise} sample and uniformly distributed over the sky. We plan to work on such a selection in the near future.

\subsection{Anomalies with intermediate $W1-W2$ colour: mixture of stars and compact galaxies}

\begin{figure}
  \centering
\includegraphics[width=.95\linewidth]{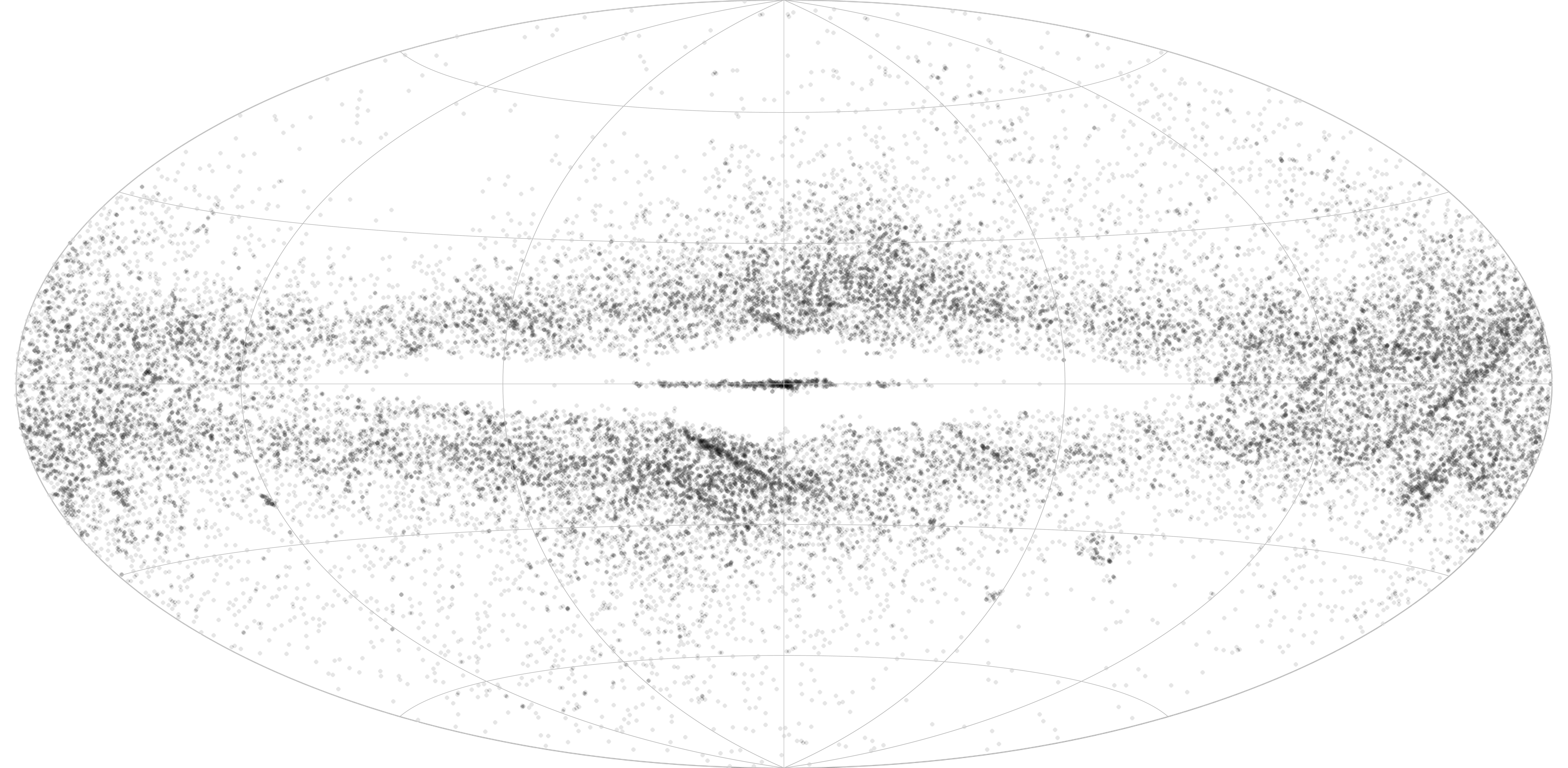}
  \caption{Sky distribution of AllWISE anomalous sources with $0\le W1-W2<0.8$ (26~990 objects).}
\label{fig:sky_centralcloud}
\end{figure}

We find approximately 27~000 anomalous sources with intermediate $W1-W2$ colours ($0\le W1-W2<0.8$), mostly located at low Galactic latitudes but outside the Bulge area, except for a small fraction by the Galactic Centre which again are supposedly photometric artefacts (Fig.~\ref{fig:sky_centralcloud}). Interestingly, there is an enhancement in surface density of these outliers also by the Galactic {Anti}centre. 
Only 14\% of them are located at $|b_\mathrm{Gal}|>30^\circ$, i.e. on half of the sky. Regarding their photometric properties in WISE, these sources are mostly faint, peaking at the limit of the catalogue, $W1=16$. About half of them have $W3$ detections, which is very different from the AGN candidate case from the previous section. 
Moreover, they appear to be very compact, having \texttt{w1mag13} values below $0.1$. This  causes their anomalous behaviour for the algorithm, as practically no sources in the training sample have this property (cf. Fig.~\ref{fig:w13}).

 Similarly to what was done in Sec.~\ref{sec:QSOs}, we paired up this sample with external datasets. Unlike in that case, here almost exactly a half of them have counterparts in 2MASS PSC, and the sky distribution of the matches roughly follows that of this anomaly sample. Except for a handful of real artefacts from the Galactic Bulge, all these objects are faint in the $J H K_s$ bands.
 On the other hand, exactly none (0) of these anomalies have a counterpart in 2MASS XSC.  Altogether, this means that except for obvious artefacts, these sources are either stars or compact galaxies unresolved by 2MASS.

\begin{figure}
  \centering
\includegraphics[width=.9\linewidth]{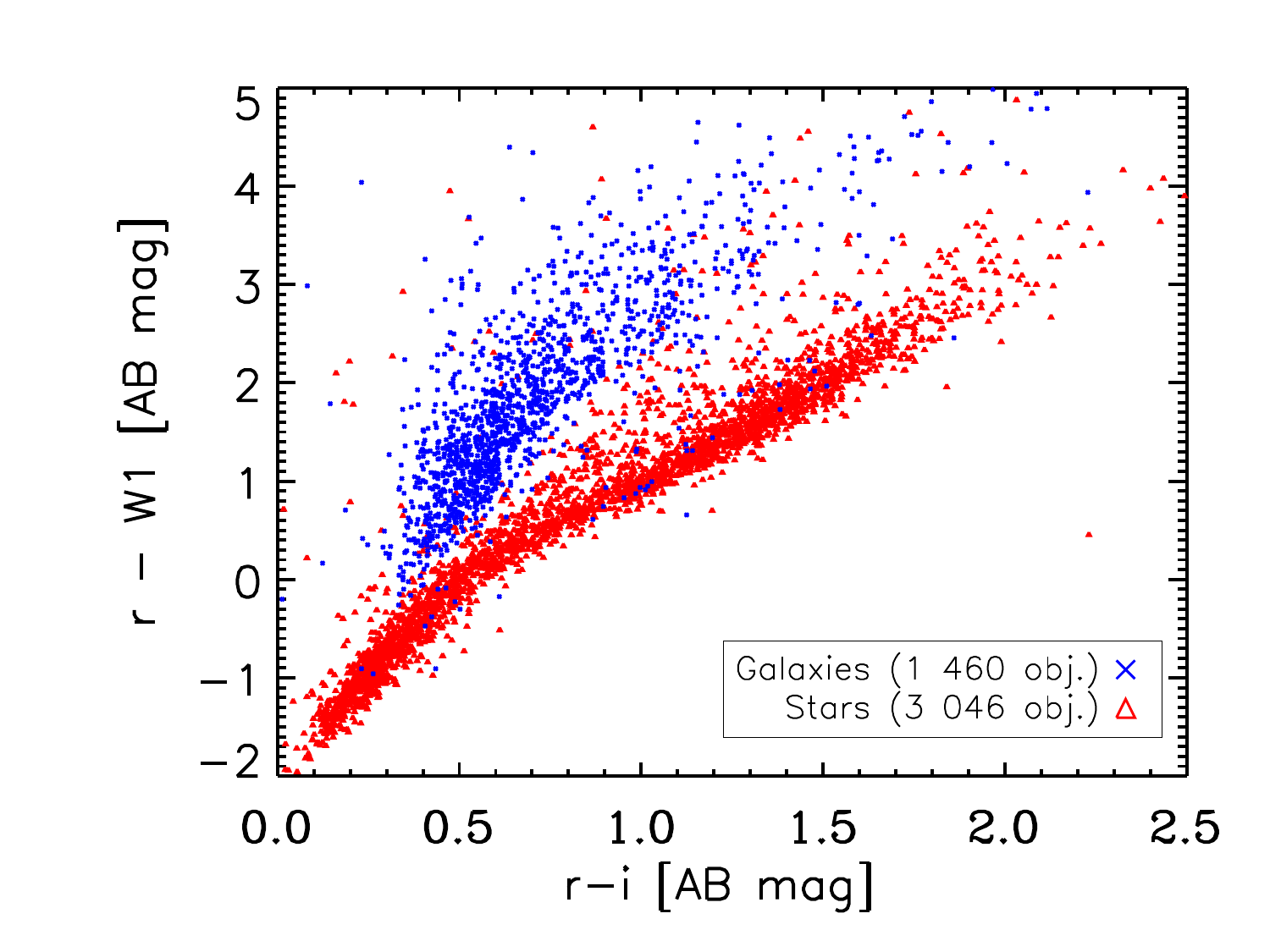}
  \caption{Optical-infrared colour-colour diagram for sources in the AllWISE anomaly sample with intermediate $0\le W1-W2<0.8$ and also present  in the SDSS photometric dataset. Blue and red dots are respectively stars and galaxies according to SDSS morphological classification. All magnitudes are here AB; WISE $W1$ was converted following \cite{jarrett11}.}
  \label{fig:rirw1}
\end{figure}

Further evidence that they are a mixture of these two source types comes from a cross-match with the SDSS photometric catalogue.  Here we find only  $\sim4500$ matches, partly because this outlier dataset overlaps with SDSS footprint only to small extent, with practically no anomalies in the north Galactic cap where the SDSS coverage is the best. As shown in \cite{Prakash15}, a combination of optical $r i$ bands and WISE $W1$ can be used for an efficient separation of stars from galaxies
(see also \citealt{DESI} for a similar separation using the $z$ band rather than $i$). We indeed observe such a division in our anomaly sample, as shown in Fig.~\ref{fig:rirw1}; magnitudes are in the AB system for a straightforward comparison with the SDSS and DESI studies. Colour-coding by SDSS morphological classification (blue=stars; red=galaxies) confirms the two-class nature of the part of the anomaly sample which has SDSS counterparts. The galaxies from this anomaly subset matched with SDSS (1~500) are also present  in the SDSS DR12 photometric redshift catalogue \citep{beck16} and their redshift distribution is slightly shifted towards smaller redshifts than the SDSS DR12 sample. 

Based on the above considerations, we conclude that the AllWISE anomalies identified by OCSVM with $0\le W1-W2<0.8$ colour are a mixture of stars (probably dominating), compact galaxies outside the local volume, and a handful of actual artefacts. However, only with the addition of optical photometry does the distinction between stars and galaxies become more straightforward.
To distinguish between stars and galaxies without resorting to optical measurements, i.e. based on WISE data alone, proper motion measurements could possibly be used. We note that \cite{kurcz16}  made an attempt to use proper motions as a discriminating parameter for automated source classification in AllWISE data, but no improvement was found after adding them, and actually the accuracy of the classifier was decreasing. This effect can be attributed to the fact that the WISE proper motions are not yet accurate enough to be used in source classification; they are reliable only for a small subset of WISE with high signal-to-noise ratio \citep{kirk14,kirk16}. In future WISE data releases \citep{faherty15,meisner17a,meisner17b} and in planned  surveys like LSST, proper motions should become an important parameter to be used in classification schemes.

\section{Summary and future prospects }
\label{sec:Summary}
In this work we demonstrated the power of automatic semi-supervised outlier detection, based on one-class reformulation of the SVM algorithm, and applied it to the WISE survey.
By design the algorithm creates a model of standard patterns and relations in the data through training on a set of known objects.
Then, data from a target set can be fitted to the model of normality and classified as either {known} or {unknown}.

The most relevant feature  of the OCSVM algorithm is its ability to detect real outliers among the sources in the given dataset. In the present application to AllWISE, we found three main groups of such anomalies. The first group contains  actual photometric artefacts located in areas of high surface density which have underestimated fluxes at 4.6 $\mu$m, most likely due to blending, and thus unphysical mid-IR colours. The second group includes real astrophysical objects, whose nature is consistent  with that of a dusty AGN population. Their main outlying property is their very red mid-IR colour, which caused these sources to become unclassifiable in the classical approaches to automated AllWISE object divisions. 
 The third group of anomalous sources are a mix of IR-bright stars and compact galaxies, underrepresented in the optically selected training set.

By adding these so far missing sources specific to mid-IR selection but not present in optically driven training sets, automated source separation in AllWISE data should provide more reliable results that  was possible with the classical SVM approach presented in \cite{kurcz16}. 
For the best performance of the SVM-based automated source classification it would be advisable to apply a novelty detection on the data before the traditional classification is conducted. This approach should ensure that the training sample includes a sufficient number of  object types contained within the survey and should bring insight into how well the training sample built from another survey can represent the data to be classified.

 The OCSVM algorithm can be used not only as an outlier detector, but also as a means of testing the adequacy of training samples for fully supervised classification methods. 
 When a pattern in the data does not match any of the previously learned templates, a standard supervised classifier will assign the membership to a randomly chosen class. The versatility of the OCSVM algorithm to deal with outlying and otherwise unclassifiable data should be taken advantage of by using it as a primary step to create reliable training samples as well as to provide insight into what types of objects the supervisor {should} expect to find.

In a more remote future, algorithms such as OCSVM should prove essential in the efficient search for novel, unexpected, or just rare objects in the ever growing volume of data collected by planned surveys like SPICA, SKA, or LSST.

\begin{acknowledgements}
This publication makes use of data products from the Wide-field Infrared Survey Explorer, which is a joint project of the University of California, Los Angeles, and the Jet Propulsion Laboratory/California Institute of Technology, funded by the National Aeronautics and Space Administration.
Funding for SDSS-III has been provided by the Alfred P. Sloan Foundation, the Participating Institutions, the National Science Foundation, and the U.S. Department of Energy Office of Science. The SDSS-III web site is http://www.sdss3.org/.
SDSS-III is managed by the Astrophysical Research Consortium for the Participating Institutions of the SDSS-III Collaboration including the University of Arizona, the Brazilian Participation Group, Brookhaven National Laboratory, Carnegie Mellon University, University of Florida, the French Participation Group, the German Participation Group, Harvard University, the Instituto de Astrofisica de Canarias, the Michigan State/Notre Dame/JINA Participation Group, Johns Hopkins University, Lawrence Berkeley National Laboratory, Max Planck Institute for Astrophysics, Max Planck Institute for Extraterrestrial Physics, New Mexico State University, New York University, Ohio State University, Pennsylvania State University, University of Portsmouth, Princeton University, the Spanish Participation Group, University of Tokyo, University of Utah, Vanderbilt University, University of Virginia, University of Washington, and Yale University.
The authors would like to thank the anonymous referee for the comments and recommendations which helped to improve this manuscript.
Special thanks to Mark Taylor for the TOPCAT \citep{TOPCAT} and STILTS \citep{STILTS} software.
This research has made use of Aladin sky atlas developed at CDS, Strasbourg Observatory, France.
This research has been supported by National  Science  Centre grants number UMO-2015/16/S/ST9/00438, UMO-2012/07/D/ST9/02785, UMO-2012/07/B/ST9/04425, and UMO-2015/17/D/ST9/02121.

\end{acknowledgements}

\bibliographystyle{aa}
\bibliography{ocsvm}

\begin{thebibliography}{92}
\expandafter\ifx\csname natexlab\endcsname\relax\def\natexlab#1{#1}\fi

\bibitem[{Agyemang {et~al.}(2006)Agyemang, Barker, \& Alhajj}]{agyemang06}
Agyemang, M., Barker, K., \& Alhajj, R. 2006, Intell. Data Anal., 10, 521

\bibitem[{Angiulli {et~al.}(2009)Angiulli, Fassetti, \& Palopoli}]{angiulli09}
Angiulli, F., Fassetti, F., \& Palopoli, L. 2009, ACM Trans. Database Syst.,
  34, 7:1

\bibitem[{{Banerji} {et~al.}(2013){Banerji}, {McMahon}, {Hewett},
  {Gonzalez-Solares}, \& {Koposov}}]{banerji13}
{Banerji}, M., {McMahon}, R.~G., {Hewett}, P.~C., {Gonzalez-Solares}, E., \&
  {Koposov}, S.~E. 2013, \mnras, 429, L55

\bibitem[{{Baron} \& {Poznanski}(2017)}]{baron16}
{Baron}, D. \& {Poznanski}, D. 2017, \mnras, 465, 4530

\bibitem[{Basu {et~al.}(2004)Basu, Bilenko, \& Mooney}]{basu04}
Basu, S., Bilenko, M., \& Mooney, R.~J. 2004, in Proceedings of the Tenth ACM
  SIGKDD International Conference on Knowledge Discovery and Data Mining, KDD
  '04 (New York, NY, USA: ACM), 59--68

\bibitem[{Batuwita \& Palade(2013)}]{imbalance}
Batuwita, R. \& Palade, V. 2013, Class Imbalance Learning Methods for Support
  Vector Machines (John Wiley and Sons, Inc.), 83--99

\bibitem[{{Beaumont} {et~al.}(2011){Beaumont}, {Williams}, \&
  {Goodman}}]{beaumont2011}
{Beaumont}, C.~N., {Williams}, J.~P., \& {Goodman}, A.~A. 2011, \apj, 741, 14

\bibitem[{{Beck} {et~al.}(2016){Beck}, {Dobos}, {Budav{\'a}ri}, {Szalay}, \&
  {Csabai}}]{beck16}
{Beck}, R., {Dobos}, L., {Budav{\'a}ri}, T., {Szalay}, A.~S., \& {Csabai}, I.
  2016, \mnras, 460, 1371

\bibitem[{{Benjamin} {et~al.}(2003){Benjamin}, {Churchwell}, {Babler}, {Bania},
  {Clemens}, {Cohen}, {Dickey}, {Indebetouw}, {Jackson}, {Kobulnicky},
  {Lazarian}, {Marston}, {Mathis}, {Meade}, {Seager}, {Stolovy}, {Watson},
  {Whitney}, {Wolff}, \& {Wolfire}}]{glimpse}
{Benjamin}, R.~A., {Churchwell}, E., {Babler}, B.~L., {et~al.} 2003, \pasp,
  115, 953

\bibitem[{{Bilicki} {et~al.}(2014){Bilicki}, {Jarrett}, {Peacock}, {Cluver}, \&
  {Steward}}]{2MPZ}
{Bilicki}, M., {Jarrett}, T.~H., {Peacock}, J.~A., {Cluver}, M.~E., \&
  {Steward}, L. 2014, \apjs, 210, 9

\bibitem[{{Bilicki} {et~al.}(2016){Bilicki}, {Peacock}, {Jarrett}, {Cluver},
  {Maddox}, {Brown}, {Taylor}, {Hambly}, {Solarz}, {Holwerda}, {Baldry},
  {Loveday}, {Moffett}, {Hopkins}, {Driver}, {Alpaslan}, \&
  {Bland-Hawthorn}}]{bilicki16}
{Bilicki}, M., {Peacock}, J.~A., {Jarrett}, T.~H., {et~al.} 2016, \apjs, 225, 5

\bibitem[{{Blanton} \& {Roweis}(2007)}]{blanton07}
{Blanton}, M.~R. \& {Roweis}, S. 2007, \aj, 133, 734

\bibitem[{{Bolton} {et~al.}(2012){Bolton}, {Schlegel}, {Aubourg}, {Bailey},
  {Bhardwaj}, {Brownstein}, {Burles}, {Chen}, {Dawson}, {Eisenstein}, {Gunn},
  {Knapp}, {Loomis}, {Lupton}, {Maraston}, {Muna}, {Myers}, {Olmstead},
  {Padmanabhan}, {P{\^a}ris}, {Percival}, {Petitjean}, {Rockosi}, {Ross},
  {Schneider}, {Shu}, {Strauss}, {Thomas}, {Tremonti}, {Wake}, {Weaver}, \&
  {Wood-Vasey}}]{bolton12}
{Bolton}, A.~S., {Schlegel}, D.~J., {Aubourg}, {\'E}., {et~al.} 2012, \aj, 144,
  144

\bibitem[{{Brandl} {et~al.}(2006){Brandl}, {Bernard-Salas}, {Spoon}, {Devost},
  {Sloan}, {Guilles}, {Wu}, {Houck}, {Weedman}, {Armus}, {Appleton}, {Soifer},
  {Charmandaris}, {Hao}, {Higdon}, {Marshall}, \& {Herter}}]{pah}
{Brandl}, B.~R., {Bernard-Salas}, J., {Spoon}, H.~W.~W., {et~al.} 2006, \apj,
  653, 1129

\bibitem[{{Cavuoti} {et~al.}(2014){Cavuoti}, {Brescia}, {D'Abrusco}, {Longo},
  \& {Paolillo}}]{cavuoti14}
{Cavuoti}, S., {Brescia}, M., {D'Abrusco}, R., {Longo}, G., \& {Paolillo}, M.
  2014, \mnras, 437, 968

\bibitem[{{Chambers} {et~al.}(2016){Chambers}, {Magnier}, {Metcalfe},
  {Flewelling}, {Huber}, {Waters}, {Denneau}, {Draper}, {Farrow}, {Finkbeiner},
  {Holmberg}, {Koppenhoefer}, {Price}, {Saglia}, {Schlafly}, {Smartt},
  {Sweeney}, {Wainscoat}, {Burgett}, {Grav}, {Heasley}, {Hodapp}, {Jedicke},
  {Kaiser}, {Kudritzki}, {Luppino}, {Lupton}, {Monet}, {Morgan}, {Onaka},
  {Stubbs}, {Tonry}, {Banados}, {Bell}, {Bender}, {Bernard}, {Botticella},
  {Casertano}, {Chastel}, {Chen}, {Chen}, {Cole}, {Deacon}, {Frenk},
  {Fitzsimmons}, {Gezari}, {Goessl}, {Goggia}, {Goldman}, {Grebel}, {Hambly},
  {Hasinger}, {Heavens}, {Heckman}, {Henderson}, {Henning}, {Holman}, {Hopp},
  {Ip}, {Isani}, {Keyes}, {Koekemoer}, {Kotak}, {Long}, {Lucey}, {Liu},
  {Martin}, {McLean}, {Morganson}, {Murphy}, {Nieto-Santisteban}, {Norberg},
  {Peacock}, {Pier}, {Postman}, {Primak}, {Rae}, {Rest}, {Riess}, {Riffeser},
  {Rix}, {Roser}, {Schilbach}, {Schultz}, {Scolnic}, {Szalay}, {Seitz},
  {Shiao}, {Small}, {Smith}, {Soderblom}, {Taylor}, {Thakar}, {Thiel},
  {Thilker}, {Urata}, {Valenti}, {Walter}, {Watters}, {Werner}, {White},
  {Wood-Vasey}, \& {Wyse}}]{chambers}
{Chambers}, K.~C., {Magnier}, E.~A., {Metcalfe}, N., {et~al.} 2016, ArXiv
  e-prints:1612.05560

\bibitem[{Chandola {et~al.}(2009)Chandola, Banerjee, \& Kumar}]{chandola09}
Chandola, V., Banerjee, A., \& Kumar, V. 2009, ACM Comput. Surv., 41, 15:1

\bibitem[{Chapelle {et~al.}(2006)Chapelle, Sch{\"o}lkopf, \& Zien}]{chapelle06}
Chapelle, O., Sch{\"o}lkopf, B., \& Zien, A. 2006, Semi-Supervised Learning,
  Adaptive computation and machine learning (Cambridge, MA, USA: MIT Press),
  508

\bibitem[{Chapelle \& Zien(2005)}]{chapelle05}
Chapelle, O. \& Zien, A. 2005, in AISTATS 2005, Max-Planck-Gesellschaft, 57--64

\bibitem[{{Cluver} {et~al.}(2014){Cluver}, {Jarrett}, {Hopkins}, {Driver},
  {Liske}, {Gunawardhana}, {Taylor}, {Robotham}, {Alpaslan}, {Baldry}, {Brown},
  {Peacock}, {Popescu}, {Tuffs}, {Bauer}, {Bland-Hawthorn}, {Colless},
  {Holwerda}, {Lara-L{\'o}pez}, {Leschinski}, {L{\'o}pez-S{\'a}nchez},
  {Norberg}, {Owers}, {Wang}, \& {Wilkins}}]{cluver14}
{Cluver}, M.~E., {Jarrett}, T.~H., {Hopkins}, A.~M., {et~al.} 2014, \apj, 782,
  90

\bibitem[{Cortes \& Vapnik(1995)}]{cortes95}
Cortes, C. \& Vapnik, V. 1995, Mach. Learn., 20, 273

\bibitem[{{Cutri} {et~al.}(2013){Cutri}, {Wright}, {Conrow}, {Fowler},
  {Eisenhardt}, {Grillmair}, {Kirkpatrick}, {Masci}, {McCallon}, {Wheelock},
  {Fajardo-Acosta}, {Yan}, {Benford}, {Harbut}, {Jarrett}, {Lake}, {Leisawitz},
  {Ressler}, {Stanford}, {Tsai}, {Liu}, {Helou}, {Mainzer}, {Gettings},
  {Gonzalez}, {Hoffman}, {Marsh}, {Padgett}, {Skrutskie}, {Beck}, {Papin}, \&
  {Wittman}}]{cutri13}
{Cutri}, R.~M., {Wright}, E.~L., {Conrow}, T., {et~al.} 2013, {Explanatory
  Supplement to the AllWISE Data Release Products}, Tech. rep., Explanatory
  Supplement to the AllWISE Data Release Products, by R. M. Cutri et al.

\bibitem[{{DESI Collaboration} {et~al.}(2016){DESI Collaboration}, {Aghamousa},
  {Aguilar}, {Ahlen}, {Alam}, {Allen}, {Allende Prieto}, {Annis}, {Bailey},
  {Balland}, {Ballester}, {Baltay}, {Beaufore}, {Bebek}, {Beers}, {Bell},
  {Bernal}, {Besuner}, {Beutler}, {Blake}, \& et~al.}]{DESI}
{DESI Collaboration}, {Aghamousa}, A., {Aguilar}, J., {et~al.} 2016, ArXiv
  e-prints:1611.00036

\bibitem[{{Donoso} {et~al.}(2014){Donoso}, {Yan}, {Stern}, \&
  {Assef}}]{donoso14}
{Donoso}, E., {Yan}, L., {Stern}, D., \& {Assef}, R.~J. 2014, \apj, 789, 44

\bibitem[{{Fadely} {et~al.}(2012){Fadely}, {Hogg}, \& {Willman}}]{fadely2012}
{Fadely}, R., {Hogg}, D.~W., \& {Willman}, B. 2012, \apj, 760, 15

\bibitem[{{Faherty} {et~al.}(2015){Faherty}, {Alatalo}, {Anderson}, {Assef},
  {Bardalez Gagliuffi}, {Barry}, {Benford}, {Bilicki}, {Burningham},
  {Christian}, {Cushing}, {Eisenhardt}, {Elvisx}, {Fajardo-Acosta},
  {Finkbeiner}, {Fischer}, {Forrest}, {Fowler}, {Gardner}, {Gelino}, {Gorjian},
  {Grillmair}, {Gromadzki}, {Hall}, {Ivezi'c}, {Izumi}, {Kirkpatrick},
  {Kov{\'a}cs}, {Lang}, {Leisawitz}, {Liu}, {Mainzer}, {Malek}, {Marton},
  {Masci}, {McLean}, {Meisner}, {Nikutta}, {Padgett}, {Patel}, {Rebull},
  {Rich}, {Ringwald}, {Rose}, {Schneider}, {Stassun}, {Stern}, {Tsai}, {Wang},
  {Weston}, {L.}, {Wright}, {Wu}, \& {Yang}}]{faherty15}
{Faherty}, J.~K., {Alatalo}, K., {Anderson}, L.~D., {et~al.} 2015, ArXiv
  e-prints:1505.01923

\bibitem[{{Hambly} {et~al.}(2001){Hambly}, {MacGillivray}, {Read}, {Tritton},
  {Thomson}, {Kelly}, {Morgan}, {Smith}, {Driver}, {Williamson}, {Parker},
  {Hawkins}, {Williams}, \& {Lawrence}}]{scosmos}
{Hambly}, N.~C., {MacGillivray}, H.~T., {Read}, M.~A., {et~al.} 2001, \mnras,
  326, 1279

\bibitem[{Han {et~al.}(2011)Han, Kamber, \& Pei}]{clust}
Han, J., Kamber, M., \& Pei, J. 2011, Data Mining: Concepts and Techniques, 3rd
  edn. (San Francisco, CA, USA: Morgan Kaufmann Publishers Inc.)

\bibitem[{Hautamaki {et~al.}(2004)Hautamaki, Karkkainen, \&
  Franti}]{hautamaki04}
Hautamaki, V., Karkkainen, I., \& Franti, P. 2004, in Proceedings of the
  Pattern Recognition, 17th International Conference on (ICPR'04) Volume 3 -
  Volume 03, ICPR '04 (Washington, DC, USA: IEEE Computer Society), 430--433

\bibitem[{Hawkins {et~al.}(2002)Hawkins, He, Williams, \& Baxter}]{hawkins02}
Hawkins, S., He, H., Williams, G.~J., \& Baxter, R.~A. 2002, in Proceedings of
  the 4th International Conference on Data Warehousing and Knowledge Discovery,
  DaWaK 2000 (London, UK, UK: Springer-Verlag), 170--180

\bibitem[{{Heinis} {et~al.}(2016){Heinis}, {Kumar}, {Gezari}, {Burgett},
  {Chambers}, {Draper}, {Flewelling}, {Kaiser}, {Magnier}, {Metcalfe}, \&
  {Waters}}]{heinis2016}
{Heinis}, S., {Kumar}, S., {Gezari}, S., {et~al.} 2016, \apj, 821, 86

\bibitem[{Ho(1998)}]{rf98}
Ho, T.~K. 1998, IEEE Trans. Pattern Anal. Mach. Intell., 20, 832

\bibitem[{Hodge \& Austin(2004)}]{hodge04}
Hodge, V. \& Austin, J. 2004, Artif. Intell. Rev., 22, 85

\bibitem[{Hoffmann(2007)}]{hoffmann07}
Hoffmann, H. 2007, Pattern Recogn., 40, 863

\bibitem[{{Hoyle}(2016)}]{hoyle15}
{Hoyle}, B. 2016, Astronomy and Computing, 16, 34

\bibitem[{{Jarrett} {et~al.}(2000){Jarrett}, {Chester}, {Cutri}, {Schneider},
  {Skrutskie}, \& {Huchra}}]{jarrett00}
{Jarrett}, T.~H., {Chester}, T., {Cutri}, R., {et~al.} 2000, \aj, 119, 2498

\bibitem[{{Jarrett} {et~al.}(2017){Jarrett}, {Cluver}, {Magoulas}, {Bilicki},
  {Alpaslan}, {Bland-Hawthorn}, {Brough}, {Brown}, {Croom}, {Driver},
  {Holwerda}, {Hopkins}, {Loveday}, {Norberg}, {Peacock}, {Popescu}, {Sadler},
  {Taylor}, {Tuffs}, \& {Wang}}]{jarrett17}
{Jarrett}, T.~H., {Cluver}, M.~E., {Magoulas}, C., {et~al.} 2017, \apj, 836,
  182

\bibitem[{{Jarrett} {et~al.}(2011){Jarrett}, {Cohen}, {Masci}, {Wright},
  {Stern}, {Benford}, {Blain}, {Carey}, {Cutri}, {Eisenhardt}, {Lonsdale},
  {Mainzer}, {Marsh}, {Padgett}, {Petty}, {Ressler}, {Skrutskie}, {Stanford},
  {Surace}, {Tsai}, {Wheelock}, \& {Yan}}]{jarrett11}
{Jarrett}, T.~H., {Cohen}, M., {Masci}, F., {et~al.} 2011, \apj, 735, 112

\bibitem[{{Jolliffe}(2002)}]{jolliffe02}
{Jolliffe}, I. 2002, Principal component analysis (New York: Springer Verlag)

\bibitem[{{Kirkpatrick} {et~al.}(2016){Kirkpatrick}, {Kellogg}, {Schneider},
  {Fajardo-Acosta}, {Cushing}, {Greco}, {Mace}, {Gelino}, {Wright},
  {Eisenhardt}, {Stern}, {Faherty}, {Sheppard}, {Lansbury}, {Logsdon},
  {Martin}, {McLean}, {Schurr}, {Cutri}, \& {Conrow}}]{kirk16}
{Kirkpatrick}, J.~D., {Kellogg}, K., {Schneider}, A.~C., {et~al.} 2016, \apjs,
  224, 36

\bibitem[{{Kirkpatrick} {et~al.}(2014){Kirkpatrick}, {Schneider},
  {Fajardo-Acosta}, {Gelino}, {Mace}, {Wright}, {Logsdon}, {McLean}, {Cushing},
  {Skrutskie}, {Eisenhardt}, {Stern}, {Balokovi{\'c}}, {Burgasser}, {Faherty},
  {Lansbury}, {Rich}, {Skrzypek}, {Fowler}, {Cutri}, {Masci}, {Conrow},
  {Grillmair}, {McCallon}, {Beichman}, \& {Marsh}}]{kirk14}
{Kirkpatrick}, J.~D., {Schneider}, A., {Fajardo-Acosta}, S., {et~al.} 2014,
  \apj, 783, 122

\bibitem[{{Kov{\'a}cs} \& {Szapudi}(2015)}]{KoSz15}
{Kov{\'a}cs}, A. \& {Szapudi}, I. 2015, \mnras, 448, 1305

\bibitem[{{Krakowski} {et~al.}(2016){Krakowski}, {Ma{\l}ek}, {Bilicki},
  {Pollo}, {Kurcz}, \& {Krupa}}]{krakowski16}
{Krakowski}, T., {Ma{\l}ek}, K., {Bilicki}, M., {et~al.} 2016, \aap, 596, A39

\bibitem[{Kriegel {et~al.}(2009)Kriegel, Kr\"{o}ger, \& Zimek}]{kriegel09}
Kriegel, H.-P., Kr\"{o}ger, P., \& Zimek, A. 2009, ACM Trans. Knowl. Discov.
  Data, 3, 1:1

\bibitem[{{Kurcz} {et~al.}(2016){Kurcz}, {Bilicki}, {Solarz}, {Krupa}, {Pollo},
  \& {Ma{\l}ek}}]{kurcz16}
{Kurcz}, A., {Bilicki}, M., {Solarz}, A., {et~al.} 2016, \aap, 592, A25

\bibitem[{Langone {et~al.}(2015)Langone, Mall, Alzate, \& Suykens}]{langone15}
Langone, R., Mall, R., Alzate, C., \& Suykens, J. A.~K. 2015, CoRR,
  abs/1505.00477

\bibitem[{Le {et~al.}(2010)Le, Tran, Ma, \& Sharma}]{TrungLe10}
Le, T., Tran, D., Ma, W., \& Sharma, D. 2010, An optimal sphere and two large
  margins approach for novelty detection

\bibitem[{Le {et~al.}(2011)Le, Tran, Ma, \& Sharma}]{Le11}
Le, T., Tran, D., Ma, W., \& Sharma, D. 2011, Multiple Distribution Data
  Description Learning Algorithm for Novelty Detection

\bibitem[{Liu {et~al.}(2011)Liu, Liu, \& Chen}]{Liu11}
Liu, Y.-H., Liu, Y.-C., \& Chen, Y.-Z. 2011, Expert Syst. Appl., 38, 6222

\bibitem[{{Mainzer} {et~al.}(2014){Mainzer}, {Bauer}, {Cutri}, {Grav},
  {Masiero}, {Beck}, {Clarkson}, {Conrow}, {Dailey}, {Eisenhardt}, {Fabinsky},
  {Fajardo-Acosta}, {Fowler}, {Gelino}, {Grillmair}, {Heinrichsen}, {Kendall},
  {Kirkpatrick}, {Liu}, {Masci}, {McCallon}, {Nugent}, {Papin}, {Rice},
  {Royer}, {Ryan}, {Sevilla}, {Sonnett}, {Stevenson}, {Thompson}, {Wheelock},
  {Wiemer}, {Wittman}, {Wright}, \& {Yan}}]{mainzer14}
{Mainzer}, A., {Bauer}, J., {Cutri}, R.~M., {et~al.} 2014, \apj, 792, 30

\bibitem[{{Ma{\l}ek} {et~al.}(2013){Ma{\l}ek}, {Solarz}, {Pollo}, {Fritz},
  {Garilli}, {Scodeggio}, {Iovino}, {Granett}, {Abbas}, {Adami}, {Arnouts},
  {Bel}, {Bolzonella}, {Bottini}, {Branchini}, {Cappi}, {Coupon}, {Cucciati},
  {Davidzon}, {De Lucia}, {de la Torre}, {Franzetti}, {Fumana}, {Guzzo},
  {Ilbert}, {Krywult}, {Le Brun}, {Le Fevre}, {Maccagni}, {Marulli},
  {McCracken}, {Paioro}, {Polletta}, {Schlagenhaufer}, {Tasca}, {Tojeiro},
  {Vergani}, {Zanichelli}, {Burden}, {Di Porto}, {Marchetti}, {Marinoni},
  {Mellier}, {Moscardini}, {Nichol}, {Peacock}, {Percival}, {Phleps}, {Wolk},
  \& {Zamorani}}]{malek13}
{Ma{\l}ek}, K., {Solarz}, A., {Pollo}, A., {et~al.} 2013, \aap, 557, A16

\bibitem[{Manevitz \& Yousef(2007)}]{Manevitz07}
Manevitz, L. \& Yousef, M. 2007, Neurocomput., 70, 1466

\bibitem[{Markou \& Singh(2003)}]{markou02}
Markou, M. \& Singh, S. 2003, Signal Processing, 83, 2499

\bibitem[{{Marton} {et~al.}(2016){Marton}, {T{\'o}th}, {Paladini}, {Kun},
  {Zahorecz}, {McGehee}, \& {Kiss}}]{marton2016}
{Marton}, G., {T{\'o}th}, L.~V., {Paladini}, R., {et~al.} 2016, \mnras, 458,
  3479

\bibitem[{{Mateos} {et~al.}(2012){Mateos}, {Alonso-Herrero}, {Carrera},
  {Blain}, {Watson}, {Barcons}, {Braito}, {Severgnini}, {Donley}, \&
  {Stern}}]{mateos12}
{Mateos}, S., {Alonso-Herrero}, A., {Carrera}, F.~J., {et~al.} 2012, \mnras,
  426, 3271

\bibitem[{{Meisner} {et~al.}(2017{\natexlab{a}}){Meisner}, {Lang}, \&
  {Schlegel}}]{meisner17a}
{Meisner}, A.~M., {Lang}, D., \& {Schlegel}, D.~J. 2017{\natexlab{a}}, ArXiv
  e-prints:1705.06746

\bibitem[{{Meisner} {et~al.}(2017{\natexlab{b}}){Meisner}, {Lang}, \&
  {Schlegel}}]{meisner17b}
{Meisner}, A.~M., {Lang}, D., \& {Schlegel}, D.~J. 2017{\natexlab{b}}, \aj,
  153, 38

\bibitem[{Mercer(1909)}]{Mercer415}
Mercer, J. 1909, Philosophical Transactions of the Royal Society of London A:
  Mathematical, Physical and Engineering Sciences, 209, 415

\bibitem[{Meyer {et~al.}(2015)Meyer, Dimitriadou, Hornik, Weingessel, \&
  Leisch}]{e1071}
Meyer, D., Dimitriadou, E., Hornik, K., Weingessel, A., \& Leisch, F. 2015,
  e1071: Misc Functions of the Department of Statistics, Probability Theory
  Group (Formerly: E1071), TU Wien, r package version 1.6-7

\bibitem[{Mika {et~al.}(1999)Mika, R{\"a}tsch, Weston, Sch{\"o}lkopf, \&
  M{\"u}ller}]{mika99}
Mika, S., R{\"a}tsch, G., Weston, J., Sch{\"o}lkopf, B., \& M{\"u}ller, K.-R.
  1999, in Proceedings of the 1999 IEEE Signal Processing Society Workshop,
  Vol.~9, Max-Planck-Gesellschaft (IEEE), 41--48

\bibitem[{Murphy(2012)}]{murphy12}
Murphy, K.~P. 2012, Machine Learning: A Probabilistic Perspective (The MIT
  Press)

\bibitem[{{Pollo} {et~al.}(2010){Pollo}, {Rybka}, \& {Takeuchi}}]{pollo10}
{Pollo}, A., {Rybka}, P., \& {Takeuchi}, T.~T. 2010, \aap, 514, A3

\bibitem[{{Prakash} {et~al.}(2015){Prakash}, {Licquia}, {Newman}, \&
  {Rao}}]{Prakash15}
{Prakash}, A., {Licquia}, T.~C., {Newman}, J.~A., \& {Rao}, S.~M. 2015, \apj,
  803, 105

\bibitem[{{R Core Team}(2013)}]{rrr}
{R Core Team}. 2013, R: A Language and Environment for Statistical Computing, R
  Foundation for Statistical Computing, Vienna, Austria

\bibitem[{{Rahman} {et~al.}(2016){Rahman}, {M{\'e}nard}, \&
  {Scranton}}]{rahman16}
{Rahman}, M., {M{\'e}nard}, B., \& {Scranton}, R. 2016, \mnras, 457, 3912

\bibitem[{Sangeetha \& Kalpana(2010)}]{Sangeetha2010}
Sangeetha, R. \& Kalpana, B. 2010, A Comparative Study and Choice of an
  Appropriate Kernel for Support Vector Machines, ed. V.~V. Das \&
  R.~Vijaykumar (Berlin, Heidelberg: Springer Berlin Heidelberg), 549--553

\bibitem[{{Sauvage} {et~al.}(2005){Sauvage}, {Tuffs}, \& {Popescu}}]{sauvage05}
{Sauvage}, M., {Tuffs}, R.~J., \& {Popescu}, C.~C. 2005, \ssr, 119, 313

\bibitem[{Sch\"{o}lkopf {et~al.}(1999)Sch\"{o}lkopf, Smola, \&
  M\"{u}ller}]{scholkopf99}
Sch\"{o}lkopf, B., Smola, A.~J., \& M\"{u}ller, K.-R. 1999, in Advances in
  Kernel Methods, ed. B.~Sch\"{o}lkopf, C.~J.~C. Burges, \& A.~J. Smola
  (Cambridge, MA, USA: MIT Press), 327--352

\bibitem[{{Sch{\"o}lkopf} {et~al.}(2000){Sch{\"o}lkopf}, {Williamson}, {Smola},
  {Shawe-Taylor}, \& {Platt}}]{scholkopf}
{Sch{\"o}lkopf}, B., {Williamson}, R., {Smola}, A., {Shawe-Taylor}, J., \&
  {Platt}, J. 2000, Adv. Neural Inf. Process. Syst., 582–588

\bibitem[{{SDSS Collaboration} {et~al.}(2016){SDSS Collaboration}, {Albareti},
  {Allende Prieto}, {Almeida}, {Anders}, {Anderson}, {Andrews},
  {Aragon-Salamanca}, {Argudo-Fernandez}, {Armengaud}, \& et~al.}]{sdssdr13}
{SDSS Collaboration}, {Albareti}, F.~D., {Allende Prieto}, C., {et~al.} 2016,
  ArXiv e-prints:1608.02013

\bibitem[{{Secrest} {et~al.}(2015){Secrest}, {Dudik}, {Dorland}, {Zacharias},
  {Makarov}, {Fey}, {Frouard}, \& {Finch}}]{agnallwise}
{Secrest}, N.~J., {Dudik}, R.~P., {Dorland}, B.~N., {et~al.} 2015, \apjs, 221,
  12

\bibitem[{{Shawe-Taylor} \& {Cristianini}(2004)}]{st}
{Shawe-Taylor}, S. \& {Cristianini}, N. 2004, Kernel Methods for Pattern
  Analysis (Cambridge, UK: Cambridge, UP)

\bibitem[{{Shi} {et~al.}(2015){Shi}, {Liu}, {Sun}, {Li}, {Lei}, \&
  {Wang}}]{shi15}
{Shi}, F., {Liu}, Y.-Y., {Sun}, G.-L., {et~al.} 2015, \mnras, 453, 122

\bibitem[{{Skrutskie} {et~al.}(2006){Skrutskie}, {Cutri}, {Stiening},
  {Weinberg}, {Schneider}, {Carpenter}, {Beichman}, {Capps}, {Chester},
  {Elias}, {Huchra}, {Liebert}, {Lonsdale}, {Monet}, {Price}, {Seitzer},
  {Jarrett}, {Kirkpatrick}, {Gizis}, {Howard}, {Evans}, {Fowler}, {Fullmer},
  {Hurt}, {Light}, {Kopan}, {Marsh}, {McCallon}, {Tam}, {Van Dyk}, \&
  {Wheelock}}]{2mass06}
{Skrutskie}, M.~F., {Cutri}, R.~M., {Stiening}, R., {et~al.} 2006, \aj, 131,
  1163

\bibitem[{{Solarz} {et~al.}(2015){Solarz}, {Pollo}, {Takeuchi}, {Ma{\l}ek},
  {Matsuhara}, {White}, {P{\c e}piak}, {Goto}, {Wada}, {Oyabu}, {Takagi},
  {Ohyama}, {Pearson}, {Hanami}, {Ishigaki}, \& {Malkan}}]{solarz2015}
{Solarz}, A., {Pollo}, A., {Takeuchi}, T.~T., {et~al.} 2015, \aap, 582, A58

\bibitem[{{Solarz} {et~al.}(2012){Solarz}, {Pollo}, {Takeuchi}, {P{\c e}piak},
  {Matsuhara}, {Wada}, {Oyabu}, {Takagi}, {Goto}, {Ohyama}, {Pearson},
  {Hanami}, \& {Ishigaki}}]{solarz2012}
{Solarz}, A., {Pollo}, A., {Takeuchi}, T.~T., {et~al.} 2012, \aap, 541, A50

\bibitem[{{Stern} {et~al.}(2012){Stern}, {Assef}, {Benford}, {Blain}, {Cutri},
  {Dey}, {Eisenhardt}, {Griffith}, {Jarrett}, {Lake}, {Masci}, {Petty},
  {Stanford}, {Tsai}, {Wright}, {Yan}, {Harrison}, \& {Madsen}}]{stern12}
{Stern}, D., {Assef}, R.~J., {Benford}, D.~J., {et~al.} 2012, \apj, 753, 30

\bibitem[{Tax \& Duin(2004)}]{Tax2004}
Tax, D.~M. \& Duin, R.~P. 2004, Machine Learning, 54, 45

\bibitem[{Tax \& Duin(1999)}]{Tax99}
Tax, D. M.~J. \& Duin, R. P.~W. 1999, Pattern Recognition Letters, 20, 1191

\bibitem[{{Taylor}(2005)}]{TOPCAT}
{Taylor}, M.~B. 2005, in Astronomical Society of the Pacific Conference Series,
  Vol. 347, Astronomical Data Analysis Software and Systems XIV, ed.
  P.~{Shopbell}, M.~{Britton}, \& R.~{Ebert}, 29

\bibitem[{{Taylor}(2006)}]{STILTS}
{Taylor}, M.~B. 2006, in Astronomical Society of the Pacific Conference Series,
  Vol. 351, Astronomical Data Analysis Software and Systems XV, ed.
  C.~{Gabriel}, C.~{Arviset}, D.~{Ponz}, \& S.~{Enrique}, 666

\bibitem[{{{\v S}koda} {et~al.}(2016{\natexlab{a}}){{\v S}koda}, {Pali{\v
  c}ka}, {Koza}, \& {Shakurova}}]{skoda16}
{{\v S}koda}, P., {Pali{\v c}ka}, A., {Koza}, J., \& {Shakurova}, K.
  2016{\natexlab{a}}, ArXiv e-prints:1612.07536

\bibitem[{{{\v S}koda} {et~al.}(2016{\natexlab{b}}){{\v S}koda}, {Shakurova},
  {Koza}, \& {Pali{\v c}ka}}]{skoda162}
{{\v S}koda}, P., {Shakurova}, K., {Koza}, J., \& {Pali{\v c}ka}, A.
  2016{\natexlab{b}}, ArXiv e-prints:1612.07549

\bibitem[{Vapnik \& Chervonenkis(1974)}]{vapnik74}
Vapnik, V. \& Chervonenkis, A. 1974, Theory of Pattern Recognition [in Russian]
  (Moscow: Nauka), (German Translation: W.~Wapnik \& A.~Tscherwonenkis, {\em
  Theorie der Zeichenerkennung}, Akademie--Verlag, Berlin, 1979)

\bibitem[{Vapnik(1995)}]{vapnik}
Vapnik, V.~N. 1995, The nature of statistical learning theory (New York, NY,
  USA: Springer-Verlag New York, Inc.)

\bibitem[{{Walker} {et~al.}(1989){Walker}, {Volk}, {Wainscoat}, {Schwartz}, \&
  {Cohen}}]{walker89}
{Walker}, H.~J., {Volk}, K., {Wainscoat}, R.~J., {Schwartz}, D.~E., \& {Cohen},
  M. 1989, \aj, 98, 2163

\bibitem[{{Wolf} {et~al.}(2001){Wolf}, {Meisenheimer}, {R{\"o}ser}, {Beckwith},
  {Chaffee}, {Fried}, {Hippelein}, {Huang}, {K{\"u}mmel}, {von Kuhlmann},
  {Maier}, {Phleps}, {Rix}, {Thommes}, \& {Thompson}}]{wolf01}
{Wolf}, C., {Meisenheimer}, K., {R{\"o}ser}, H.-J., {et~al.} 2001, \aap, 365,
  681

\bibitem[{{Wright} {et~al.}(2010){Wright}, {Eisenhardt}, {Mainzer}, {Ressler},
  {Cutri}, {Jarrett}, {Kirkpatrick}, {Padgett}, {McMillan}, {Skrutskie},
  {Stanford}, {Cohen}, {Walker}, {Mather}, {Leisawitz}, {Gautier}, {McLean},
  {Benford}, {Lonsdale}, {Blain}, {Mendez}, {Irace}, {Duval}, {Liu}, {Royer},
  {Heinrichsen}, {Howard}, {Shannon}, {Kendall}, {Walsh}, {Larsen}, {Cardon},
  {Schick}, {Schwalm}, {Abid}, {Fabinsky}, {Naes}, \& {Tsai}}]{wright10}
{Wright}, E.~L., {Eisenhardt}, P.~R.~M., {Mainzer}, A.~K., {et~al.} 2010, \aj,
  140, 1868

\bibitem[{{Yan} {et~al.}(2013){Yan}, {Donoso}, {Tsai}, {Stern}, {Assef},
  {Eisenhardt}, {Blain}, {Cutri}, {Jarrett}, {Stanford}, {Wright}, {Bridge}, \&
  {Riechers}}]{yan13}
{Yan}, L., {Donoso}, E., {Tsai}, C.-W., {et~al.} 2013, \aj, 145, 55

\bibitem[{Yang \& Wang(2003)}]{cluseq}
Yang, J. \& Wang, W. 2003, {Cluseq: efficient and effective sequence
  clustering}

\bibitem[{{York} {et~al.}(2000){York}, {Adelman}, {Anderson}, {Anderson},
  {Annis}, {Bahcall}, {Bakken}, {Barkhouser}, {Bastian}, {Berman}, {Boroski},
  {Bracker}, {Briegel}, {Briggs}, {Brinkmann}, {Brunner}, {Burles}, {Carey},
  {Carr}, {Castander}, {Chen}, {Colestock}, {Connolly}, {Crocker}, {Csabai},
  {Czarapata}, {Davis}, {Doi}, {Dombeck}, {Eisenstein}, {Ellman}, {Elms},
  {Evans}, {Fan}, {Federwitz}, {Fiscelli}, {Friedman}, {Frieman}, {Fukugita},
  {Gillespie}, {Gunn}, {Gurbani}, {de Haas}, {Haldeman}, {Harris}, {Hayes},
  {Heckman}, {Hennessy}, {Hindsley}, {Holm}, {Holmgren}, {Huang}, {Hull},
  {Husby}, {Ichikawa}, {Ichikawa}, {Ivezi{\'c}}, {Kent}, {Kim}, {Kinney},
  {Klaene}, {Kleinman}, {Kleinman}, {Knapp}, {Korienek}, {Kron}, {Kunszt},
  {Lamb}, {Lee}, {Leger}, {Limmongkol}, {Lindenmeyer}, {Long}, {Loomis},
  {Loveday}, {Lucinio}, {Lupton}, {MacKinnon}, {Mannery}, {Mantsch}, {Margon},
  {McGehee}, {McKay}, {Meiksin}, {Merelli}, {Monet}, {Munn}, {Narayanan},
  {Nash}, {Neilsen}, {Neswold}, {Newberg}, {Nichol}, {Nicinski}, {Nonino},
  {Okada}, {Okamura}, {Ostriker}, {Owen}, {Pauls}, {Peoples}, {Peterson},
  {Petravick}, {Pier}, {Pope}, {Pordes}, {Prosapio}, {Rechenmacher}, {Quinn},
  {Richards}, {Richmond}, {Rivetta}, {Rockosi}, {Ruthmansdorfer}, {Sandford},
  {Schlegel}, {Schneider}, {Sekiguchi}, {Sergey}, {Shimasaku}, {Siegmund},
  {Smee}, {Smith}, {Snedden}, {Stone}, {Stoughton}, {Strauss}, {Stubbs},
  {SubbaRao}, {Szalay}, {Szapudi}, {Szokoly}, {Thakar}, {Tremonti}, {Tucker},
  {Uomoto}, {Vanden Berk}, {Vogeley}, {Waddell}, {Wang}, {Watanabe},
  {Weinberg}, {Yanny}, {Yasuda}, \& {SDSS Collaboration}}]{york00}
{York}, D.~G., {Adelman}, J., {Anderson}, Jr., J.~E., {et~al.} 2000, \aj, 120,
  1579

\bibitem[{{Zhang} \& {Zhao}(2004)}]{zhang04}
{Zhang}, Y. \& {Zhao}, Y. 2004, \aap, 422, 1113

\end{thebibliography}

\end{document}